\documentclass[final,3p,times,twocolumn]{elsarticle}
\usepackage{graphicx}
\usepackage{amssymb}
\usepackage{amsmath}
\usepackage{hyperref}
\usepackage{gensymb}

\usepackage{amsmath}
\usepackage{makeidx}
 
 \journal{Physica D}

\usepackage[usenames]{color}
\usepackage[normalem]{ulem}
\usepackage{gensymb}

\usepackage{hyperref}
\usepackage[T1]{fontenc} 
\begin{document}

%{Physical Review A}

\title{Mini droplet, mega droplet and stripe formation in a dipolar condensate}

\author[fac,fac2]{Luis E. Young-S.}%\footnote{lyoung@unicartagena.edu.co}}
\ead{lyoung@unicartagena.edu.co}

 \address[fac]{Grupo de Modelado Computacional y Programa de Matem\'aticas, Facultad de Ciencias Exactas y Naturales, Universidad de Cartagena, 
130015 Cartagena de Indias, Bolivar, Colombia}
\address[fac2]{Instituto de Matem\'aticas Aplicadas, Universidad de Cartagena,
130001 Cartagena de Indias, Bolivar, Colombia}

 \author[int]{S. K. Adhikari}%\footnote{sk.adhikari@unesp.br,  professores.ift.unesp.br/sk.adhikari/}}
\ead{sk.adhikari@unesp.br}

\address[int]{Instituto de F\'{\i}sica Te\'orica, UNESP - Universidade Estadual Paulista, 01.140-070 S\~ao Paulo, S\~ao Paulo, Brazil}

\begin{abstract}

We demonstrate {\it mini} droplet, {\it mega} droplet  and {\it stripe} formation  
in a dipolar $^{164}$Dy condensate, using {an improved} mean-field model including a Lee-Huang-Yang-type interaction, employing a quasi-two-dimensional (quasi-2D) trap 
in a way distinct from that in the pioneering experiment,  M. A. Norcia et. al., Nature  596, 357 (2021), where the polarization $z$ direction was taken to be perpendicular to the quasi-2D $x$-$y$ plane. In the present study we take the polarization $z$ direction in the quasi-2D   $x$-$z$ plane.  Employing the same trapping frequencies as in the experiment, and interchanging the frequencies along the $y$ and $z$ directions, we find the formation of { mini} droplets for number of atoms $N$ as small as $N=1000$.   With the increase of number of atoms,   a
spatially-periodic supersolid-like
one-dimensional array of {mega} droplets containing $50000$ to 200000 atoms  are formed along 
the $x$ direction in the $x$-$y$ plane.  These mega droplets are elongated along the polarization $z$ direction, consequently, the spatially periodic arrangement of droplets  appears as a      
stripe pattern in the $x$-$z$ plane.  To establish the supersolidity of the droplets we 
demonstrate continued dipole-mode and scissors-mode oscillations of the droplet-lattice pattern.  
The main findings of the present study can be tested experimentally with the present know-how.  

\end{abstract}

% \pacs{03.75.Hh, 03.75.Mn, 03.75.Kk, 03.75.Lm}

\maketitle

\section{Introduction}

A quantum supersolid \cite{sprsld,sprsld1,sprsld2,sprsld3,sprsld4,sprsld5} 
exhibits a  spatially-ordered stable structure, as encountered in a 
solid crystal,  breaking continuous translational invariance and also  flows without friction  as a 
superfluid breaking  continuous gauge invariance.  Despite the failure \cite{5} of the pioneering search 
of  supersolidity in  ultracold  $^4$He in bulk \cite{4},  
the study of a supersolid has recently gained new impetus among research workers in low-temperature physics, 
after the experimental  observation of supersolids   in an SO-coupled pseudo spin-1/2 spinor Bose-Einstein condensate (BEC) 
of $^{23}$Na \cite{18} and $^{87}$Rb  \cite{donner} atoms
 as well as in a  strongly dipolar BEC.  %     \cite{2d3,drop1} of $^{164}$Dy atoms.   

In the pursuit of supersolidity,
in a quasi-one-dimensional (quasi-1D) trapped BEC of polarized dipolar atoms, a spontaneous periodic crystallization of droplets along a straight line 
was observed in different experiments  on $^{164}$Dy \cite{1d4,1d5},  $^{162}$Dy \cite{1d2,1d3,1d7}, and $^{166}$Er \cite{1d4,1d5,drop2}  atoms  after the experimental observation  of multiple droplet formation in a
quasi-1D \cite{1d1,drop1} and 
 quasi-two-dimensional
(quasi-2D) \cite{2d3}    dipolar BEC of $^{164}$Dy atoms.  More recently, supersolidity was confirmed experimentally in a quasi-2D  trapped  BEC of $^{164}$Dy atoms 
polarized along the $z$ direction through a crystallization of droplets on 
a periodic triangular lattice \cite{2d2exFer} in the $x$-$y$ plane.
A strongly dipolar BEC, with the dipolar interaction beyond a critical value, shrinks to a very small size due to an excess of dipolar attraction and eventually 
collapses in the framework of the mean-field  Gross-Pitaevskii (GP) model 
and in theoretical studies
 a higher-order Lee-Huang-Yang-type \cite{lhy} (LHY-type)     interaction \cite{qf1,qf2,qf3}
is needed      to stabilize the  strongly dipolar BEC 
 against collapse \cite{santos} and to form a droplet.
 The collapse instability of a dipolar BEC can be removed by a
  three-body repulsive interaction  \cite{7a,7b}.
   After the experimental observation  of droplet formation \cite{1d1}  and confirmation of  
 supersolidity in 
 a strongly dipolar quasi-1D \cite{1d4,1d5,1d2,1d3,1d7} and quasi-2D \cite{2d2exFer}  BECs, there have been  many theoretical studies \cite{2d4,blakieprl,other1,other2,other3,blakie,1d8,fau,kr}  on the droplet formation and supersolid crystallization of droplets in a strongly dipolar BEC. Similar crystallization on square-lattice was also demonstrated in theoretical studies \cite{luis1,box2,rev2}, in addition to the triangular-lattice crystallization of droplets in a dipolar BEC. 
 In a different setting, distinct type of droplets were found in an antidipolar BEC \cite{anti}, where by a rotating magnetic field the sign of the dipolar interaction was changed from positive to negative.
 There have been studies of dipolar supersolids in a box trap \cite{box2,box1} in a rotating trap \cite{rot}, in an infinite tube \cite{tube},  in a binary mixture \cite{bin,bin2}, and in a molecular BEC \cite{mol}.
 In a different context,  nondipolar binary  BEC droplets have been stabilized  in free space due to 
an attractive  inter-species interaction and 
a repulsive intra-species interaction  \cite{binary2,binary},  which are different from  
  the present dipolar BEC  droplets in a strong trap.
  
  %  are different from the recently observed \cite{binary,binary2}    nondipolar binary  BEC droplets in free space. Nevertheless, in both cases, the collapse is arrested by a  beyond-mean-field  LHY-type interaction.   

%\onecolumngrid

%\twocolumngrid

The droplets and droplet-lattice structure are formed for very large  atom density in a strongly 
dipolar BEC, while the system tends to collapse due to a large dipolar attraction. For a fixed number of atoms,
and atomic contact (scattering length)  and dipolar (dipole moment)  interactions, the atom density   
will increase as the harmonic trap is made stronger, e.g., for a large overall trap frequency 
$\bar \omega = \sqrt[3]{\omega_x\omega_y\omega_z}$, 
where $\omega_x,\omega_y,\omega_z$ are angular frequencies of the trap
along the $x,y,z$ directions, respectively. In  previous studies on droplet-lattice supersolids, the quasi-2D 
dipolar BEC in the $x$-$y$ plane had $\omega_z \gg \omega_x,\omega_y$.   In this paper we consider the droplet-lattice formation in a quasi-2D dipolar condensate in the  $x$-$z$ plane, {using an improved mean-field model including the LHY interaction and} 
keeping the same overall trap frequency 
$\bar \omega$
as in previous studies,  but changing the role of the angular frequencies $\omega_y$ and $\omega_z$ as 
$\omega_y \leftrightarrow \omega_z$, consequently, in this study $\omega_y \gg \omega_x,\omega_z$.
Similar trap  has previously  been used in  a dipolar BEC \cite{dipa,dipb,dipc}.
There is a dramatic change in the nature of the droplets in the new scenario. A single droplet in the present trap can be formed for as small as 1000 $^{164}$Dy  atoms, which we call a mini droplet, while in the previous trap \cite{luis1} 20000 atoms were necessary for the formation of a single droplet. For a large number of atoms,  in the new scenario a small number of droplets, each accommodating a very large number (50000 to 200000) of atoms,  which we call a mega droplet,
are formed along the $x$ axis generating a quasi-1D supersolid in the quasi-2D trap, which appear as a stripe pattern in the $x$-$z$ plane, whereas in the previous setting a large number of droplets, each containing a small number (about 10000) of atoms, are formed  in the $x$-$y$ plane resulting in a quasi-2D supersolid. 
{ Similar result has recently been confirmed in doubly dipolar   BECs \cite{chin}.  }
The formation of the quasi-1D droplet-lattice supersolid pattern along the $x$ axis  is studied via imaginary-time propagation of a mean-field model including an LHY-type interaction.

The supersolidity of the droplet-lattice pattern is established through a study of its dynamics by real-time propagation employing the converged imaginary-time wave-function as the initial state. We demonstrate continued rigid-body-like dipole-mode oscillation of the quasi-1D supersolid  without any visible distortion in a spatially-translated trap along the $x$ direction. Similar  
harmonic  scissors-mode  oscillation of the quasi-1D supersolid  in a spatially-rotated trap around the $z$ direction guarantees its supersolidity. The frequency of the dipole-mode and scissors-mode oscillations { is found to be   in good agreement} with its theoretical estimate. 
This demonstrates the dynamical stability of the crystalline structure as well as the superfluidity 
of the supersolid.

In Sec. \ref{II} we present {the improved} mean-field model including the   LHY interaction in the GP equation.  
In Sec. \ref{III} we present the numerical results for the density of  stationary  states of a single mini droplet  and of multiple mega droplets, arranged in a spatially-periodic quasi-1D lattice in a quasi-2D trap, by imaginary-time propagation.  We also report the results for  dipole-mode and scissors-mode oscillations of the quasi-1D supersolid as obtained by real-time propagation. 
    Finally, in Sec. \ref{IV} we present a summary of our findings.

\section{Improved Mean-field model}

\label{II}

At ultralow temperatures
a  BEC of $N$ dipolar atoms, polarized along the $z$ direction,
each of mass $m$   is described by the following  {3D mean-field} GP equation including the   LHY interaction for the wave function $\psi({\bf r},t)$ at time $t$ \cite{2d4,blakie,dipbec,dip,yuka,munt,munt2}
\begin{align}\label{eq.GP3d}
 i \hbar \frac{\partial \psi({\bf r},t)}{\partial t} &=\
{\Big [}  -\frac{\hbar^2}{2m}\nabla^2
+V({\bf r})
+ \frac{4\pi \hbar^2}{m}{a} N \vert \psi({\bf r},t) \vert^2 \nonumber \\
%\end{align}
%\begin{align} 
&+N \int U_{\mathrm{dd}}({\bf r-r}')  |\psi({\bf r}',t)|^2d{\bf r}'
%&\ +\frac{3\hbar^2}{m}a_{\mathrm{dd}}  N
%\int \frac{1-3\cos^2 \theta}{|{\bf R}|^3}
%\vert\psi({\mathbf r'},t)\vert^2 d{\mathbf r}'  
\nonumber \\
& +\frac{\gamma_{\mathrm{LHY}}\hbar^2}{m} N^{3/2}
|\psi({\mathbf r},t)|^3
\Big] \psi({\bf r},t), 
\end{align}
where $a$ is the atomic scattering length and 
\begin{align}
V({\bf r})&=\frac{1}{2}m(\omega_x^2x^2+\omega_y^2y^2+\omega_z ^2z^2), 
%\\
%U_{\mathrm{ext}} megastripe 
%&=\ ,
\end{align}
where $\omega_x\equiv 2\pi f_x, \omega_y \equiv 2\pi f_y, \omega_z\equiv 2\pi f_z$ are the angular frequencies of the trap
 $V({\bf r})$   along $x,y,z$ directions, respectively; the magnetic dipolar interaction between two dipolar atoms, of magnetic moment $\mu$ each, at positions $\bf r$ and $\bf r'$  is given by \cite{xx,xx2}
\begin{align}
U_{\mathrm{dd}}({\bf R}) = \frac{\mu_0 \mu^2}{4\pi} \frac{1-3\cos^2\theta }{|{\bf R}|^3}\equiv 
\frac{3\hbar^2}{m} a_{\mathrm{dd}}\frac{1-3\cos^2\theta }{|{\bf R}|^3},
\end{align}
where 
$\mu_0$ is the permeability of vacuum and $\theta$ is the angle made  by the vector $\bf R \equiv r -r'$ with the polarization $z$ direction, the dipolar length 
\begin{equation}
a_{\mathrm{dd}}\equiv \frac{\mu_0 \mu^2 m} {12\pi \hbar^2}
\end{equation}
is a measure of the strength of dipolar interaction and is used to compare the strength of dipolar interaction to contact interaction measured by the scattering length $a$.
The wave function is  normalized as $\int \vert \psi({\bf r},t) \vert^2 d{\bf r}=1$. The coefficient 
of the   LHY interaction $\gamma_{\mathrm{LHY}}$ is given by \cite{qf1,qf2,qf3,blakie}
\begin{align}\label{qf}
\gamma_{\mathrm{LHY}}= \frac{128}{3}\sqrt{\pi a^5} Q_5(\varepsilon_{\mathrm{dd}}),
\end{align}
where 
\begin{equation}
\varepsilon_{\mathrm{dd}}= \frac{a_{\mathrm{dd}}}{a}
\end{equation}
and 
the auxiliary function
\begin{equation}\label{af}
 Q_5(\varepsilon_{\mathrm{dd}})=\ \int_0^1 dx(1-\varepsilon_{\mathrm{dd}}+3x^2\varepsilon_{\mathrm{dd}})^{5/2} 
\end{equation}
 can be evaluated as \cite{blakie}
\begin{align}\label{exa}
Q_5(\varepsilon_{\mathrm{dd}}) &=\
\frac{(3\varepsilon_{\mathrm{dd}})^{5/2}}{48}  {\Re}\Big[(8+26\epsilon+33\epsilon^2)\sqrt{1+\epsilon} \nonumber\\
& + 
\ 15\epsilon^3 \mathrm{ln} \left( \frac{1+\sqrt{1+\epsilon}}{\sqrt{\epsilon}}\right)  \Big], \quad  \epsilon = \frac{1-\varepsilon_{\mathrm{dd}}}{3\varepsilon_{\mathrm{dd}}}, 
\end{align}
where $\Re$ denotes the real
 part.   
Actually, the function $Q_5$, given by Eq. (\ref{af}), as well as the coefficient $\gamma_{\mathrm{LHY}}$, representing a correction \cite{qf1,qf2,qf3} of the LHY interaction \cite{lhy} for dipolar atoms, is complex for
$ \varepsilon_{\mathrm{dd}}
>1$  and, for studies of stationary states, expression (\ref{exa}) is formally meaningful for $ \varepsilon_{\mathrm{dd}}  \le
1$ where this expression
is real \cite{qf1,qf2}. However, its imaginary part remains small compared to its real part for medium values of $a$, where
$4 \gtrapprox   \varepsilon_{\mathrm{dd}}> 1$ \cite{cnsns},  and will be neglected in this study of stationary self-bound states as in all other studies.

Equation (\ref{eq.GP3d}) can be reduced to 
the following  dimensionless form by scaling lengths in units of $l = \sqrt{\hbar/m\omega_y}$, time $t$   in units of $\omega_y^{-1}$,  energy in units of $\hbar\omega_y $, 
angular frequencies $\omega_x, \omega_y$ and $\omega_z$  of the trap in units of $\omega_y$, and density $|\psi|^2$ in units of $l^{-3}$
\begin{align}\label{GP3d}
i \frac{\partial \psi({\bf r},t)}{\partial t} &=
{\Big [}  -\frac{1}{2}\nabla^2+ V({\bf r})+ 4\pi{a} N \vert \psi({\bf r},t) \vert^2
%+U_{\mathrm{ext}}({\bf r})
%+\frac{1}{2}\Big(\frac{\omega_x^2}{\omega_y^2}x^2+ y^2+\frac{\omega_z^2}{\omega_y^2}z^2\Big)
\nonumber\\ &
+3a_{\mathrm{dd}}  N
\int \frac{1-3\cos^2 \theta}{|{\bf R}|^3}
\vert\psi({\mathbf r'},t)\vert^2 d{\mathbf r}'  \nonumber \\ &
+\gamma_{\mathrm{LHY}}N^{3/2}
|\psi({\mathbf r},t)|^3
\Big] \psi({\bf r},t),  \\
V({\bf r})&= \frac{1}{2}\Big({\omega_x^2}x^2+ y^2+{\omega_z^2}z^2\Big).
\label{GP3dx}
\end{align}
Lacking the possibility of confusion, we are using the same symbols for the dimensionless quantities. 
One can also  obtain  Eq. (\ref{GP3d})   from the variational rule
\begin{align}
i \frac{\partial \psi}{\partial t} &= \frac{\delta E}{\delta \psi^*} %, \nonumber \\
%={\Big [}  -\frac{1}{2}\nabla^2
%+U_{\mathrm{ext}}({\bf r})
%+\frac{1}{2}\Big(\frac{\omega_x^2}{\omega_z^2}x^2+ \frac{\omega_y^2}{\omega_z^2}y^2+z^2\Big)
%\nonumber\\ &+3a_{\mathrm{dd}}  N
%\int \frac{1-3\cos^2 \theta}{|{\bf R}|^3}
%\vert\psi({\mathbf r'})\vert^2 d{\mathbf r}'  \nonumber \\ &+ 4\pi{a} N \vert \psi({\bf r}) \vert^2
%+\gamma_{\mathrm{LHY}}N^{3/2}
%|\psi({\mathbf r})|^3
%\Big] \psi({\bf r}).
\end{align}
where $E$ is the energy functional  and is given by
%with the following energy functional (energy per atom)
\begin{align} \label{en}
E &= \int d{\bf r} \Big[ \frac{|\nabla\psi({\bf r})|^2}{2} +\frac{1}{2}\Big({\omega_x^2}x^2+ y^2+{\omega_z^2}z^2\Big)|\psi({\bf r})|^2\nonumber
 \\
&+ \frac{3}{2}a_{\mathrm{dd}}N|\psi({\bf r})|^2 
\left.  \int \frac{1-3\cos^2\theta}{R^3}|\psi({\bf r'})|^2 d {\bf r'} \right.  \nonumber  \\
%\end{align}
%\begin{align}
& + 2\pi Na |\psi({\bf r})|^4 +\frac{2\gamma_{\mathrm{LHY}}}{5} N^{3/2}
|\psi({\bf r})|^5\Big].
\end{align}
Expression (\ref{en}) is the energy of the BEC per atom. 
%for a stationary state.

To establish the supersolidity of the quantum states we will study their dipole-mode oscillations through real-time propagation %using the converged imaginary-time wave function as the initial state 
in the following displaced harmonic trap    
\begin{align}\label{dip}
V({\bf r})&= \frac{1}{2}\Big[{\omega_x^2}(x-x_0)^2+ y^2+{\omega_z^2}z^2\Big],
\end{align}
employing the converged imaginary-time wave function as the initial state, 
where $x_0$ is the space translation along the $x$ direction. The dipolar supersolid should  execute the simple-harmonic
oscillation $x(t) = x_0 \cos(\omega _xt)$ along the $x$ direction without any deformation   indicating
supersolidity \cite{stg}.

The angular scissors-mode oscillation of the quasi-1D supersolid is studied,  in the $x$-$y$ plane, by real-time propagation in the following space-rotated trap
\begin{align}\label{scitrap}
 U({\bf r})&={\textstyle \frac{1}{2}}\left[{\omega_x^2}(x \cos \theta_0 {+} y \sin \theta_0)^2\right.
\nonumber \\ &+ \left. ({-} x\sin \theta_0 +y \cos \theta_0 )^2+\omega_z^2 z^2\right],
\end{align}
where we again use the converged stationary-state wave function as the initial state, and  where $\theta_0$ is the angle of rotation of the potential around the polarization $z$ direction. 
In the Thomas-Fermi regime, for a large number of atoms, the BEC, obeying the  hydrodynamic equations of superfluids, 
%For a sufficiently large asymmetry of the trap in the $x$-$y$ plane a (superfluid) BEC,  { in the Thomas-Fermi regime, obeying the hydrodynamic equations of superfluids,} will 
executes the  sustained periodic scissors-mode sinusoidal  oscillation \cite{scith,sci-y} $\theta(t)=\theta_0\cos(\omega_{\mathrm{th}}t)$ with angular frequency $\omega_{\mathrm{th}}=\sqrt{(\omega_x^2+\omega_y ^2)}$.  A long-time scissors-mode oscillation without distortion ensures the rigidity of the crystalline structure as well as the 
superfluidity of the supersolid.

\section{Numerical Results}

\label{III}   

We  solve the   partial integro-differential
 equation (\ref{GP3d}) for a dipolar BEC,  numerically, by the split-time-step Crank-Nicolson
method \cite{crank}, employing the imaginary-time propagation rule, using FORTRAN/C programs \cite{dip} or their open-multiprocessing versions \cite{omp}. Due to the $1/|{\bf R}|^3$ term, it is difficult to treat numerically the nonlocal dipolar interaction integral in the GP equation  (\ref{GP3d}) including the LHY interaction in configuration space. In order to  avoid the problem, this term is evaluated in momentum space by a Fourier transformation using a convolution identity \cite{dip}, which is advantageous numerically due to the smooth behavior of this term in momentum space. The Fourier transformation of the dipolar potential in 3D is known analytically \cite{dip}, enhancing the accuracy of the numerical procedure.

In this theoretical study we will consider a dipolar BEC of $^{164}$Dy atoms and use trap frequencies as in the recent experiment on a quasi-2D supersolid \cite{2d2exFer} in the $x$-$y$ plane, where trapping frequencies along $z$ and $x$ directions  were taken as 
$f_z =167$ Hz and $f_x =  33$ Hz. To  study the transition from a quasi-1D to a quasi-2D supersolid,  the frequency  $f_y$ along $y$ direction was varied from 75 Hz to 120 Hz  in that experiment  \cite{2d2exFer}. 
%maintaining the quasi-2D condition $\omega_z \gg \omega_x, \omega_y$. 
However, in this study we consider a 
supersolid in the $x$-$z$ plane thus changing the role of $f_y$ and $f_z$ compared to that experiment \cite{2d2exFer}. Consequently, 
we take $f_y = 167$ Hz and $f_x = 33$ Hz and consider $f_z= 60$ Hz, maintaining the quasi-2D condition  in the $x$-$z$ plane
$f_y \gg f_x, f_z$,  and find that there could be a dramatic change in the formation of droplets in the quasi-2D $x$-$z$ plane compared to the droplets in the quasi-2D $x$-$y$ plane studied previously 
\cite{2d2exFer,2d4,luis1}, even if we keep the same overall frequency $\bar f =\sqrt[3]{f_x f_y f_z}$ by swapping only the $y$ and $z$ trap frequencies as $f_y \leftrightarrow  f_z$.  The BEC will be elongated along the polarization $z$ direction. If we take a smaller $f_z$ ($< 60$ Hz)  the  $z$-length will be too long 
making the numerical treatment difficult.  A larger $f_z$   ($>60$ Hz)   will  violate the required quasi-2D condition 
$f_y \gg f_x, f_z$ of the trap.

\begin{figure}[t!]
\begin{center}
\includegraphics[width=.32\linewidth]{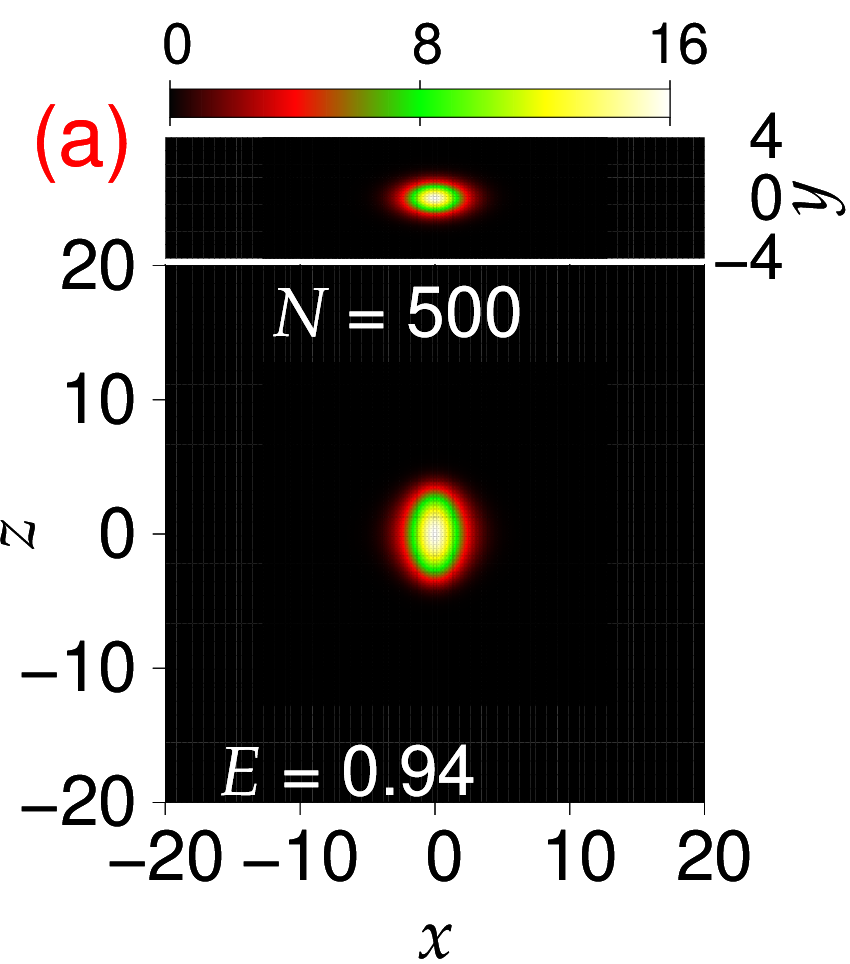}
\includegraphics[width=.32\linewidth]{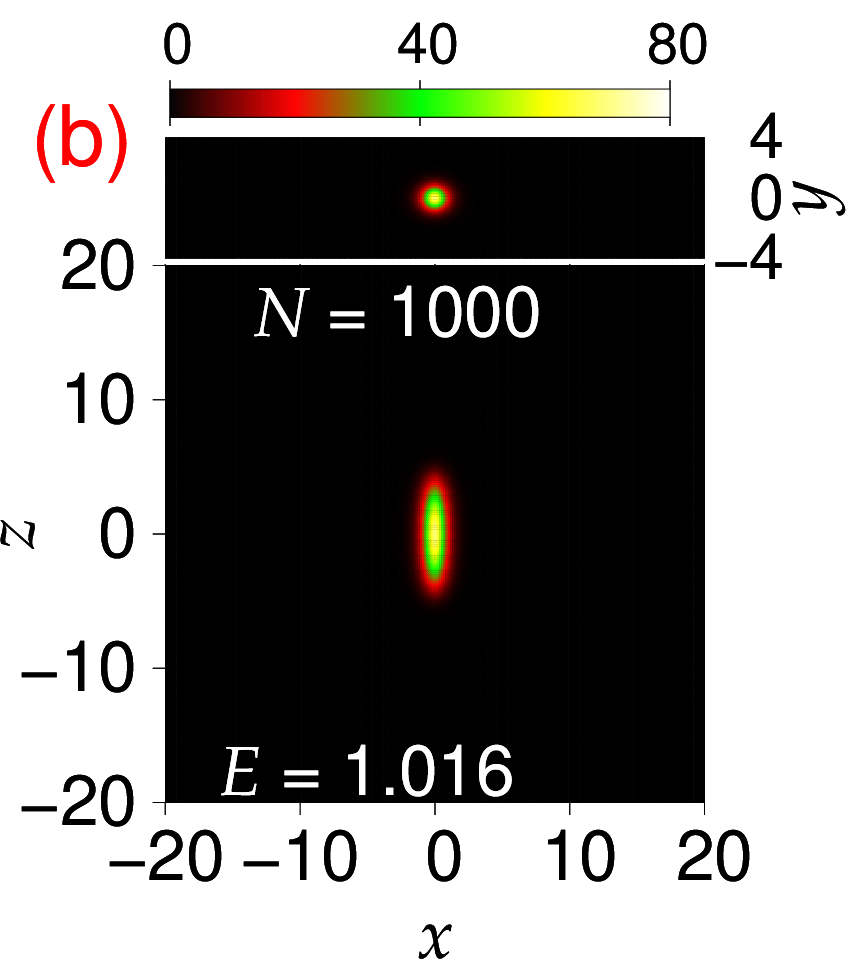}
\includegraphics[width=.32\linewidth]{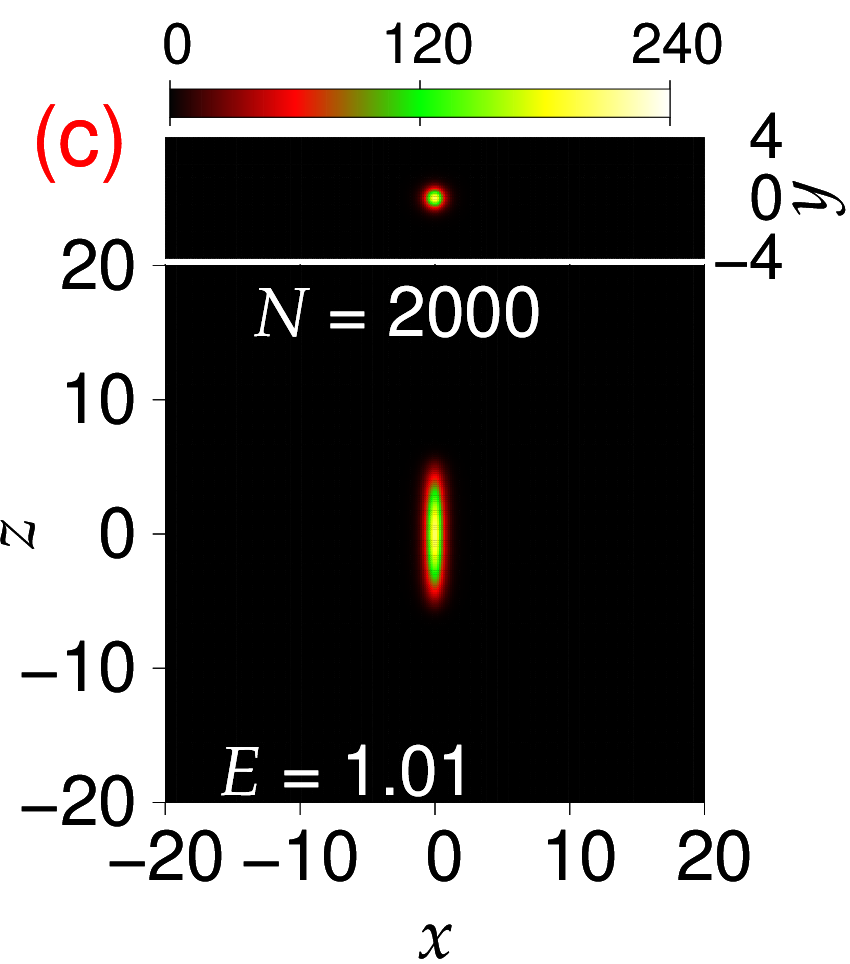}
\includegraphics[width=.32\linewidth]{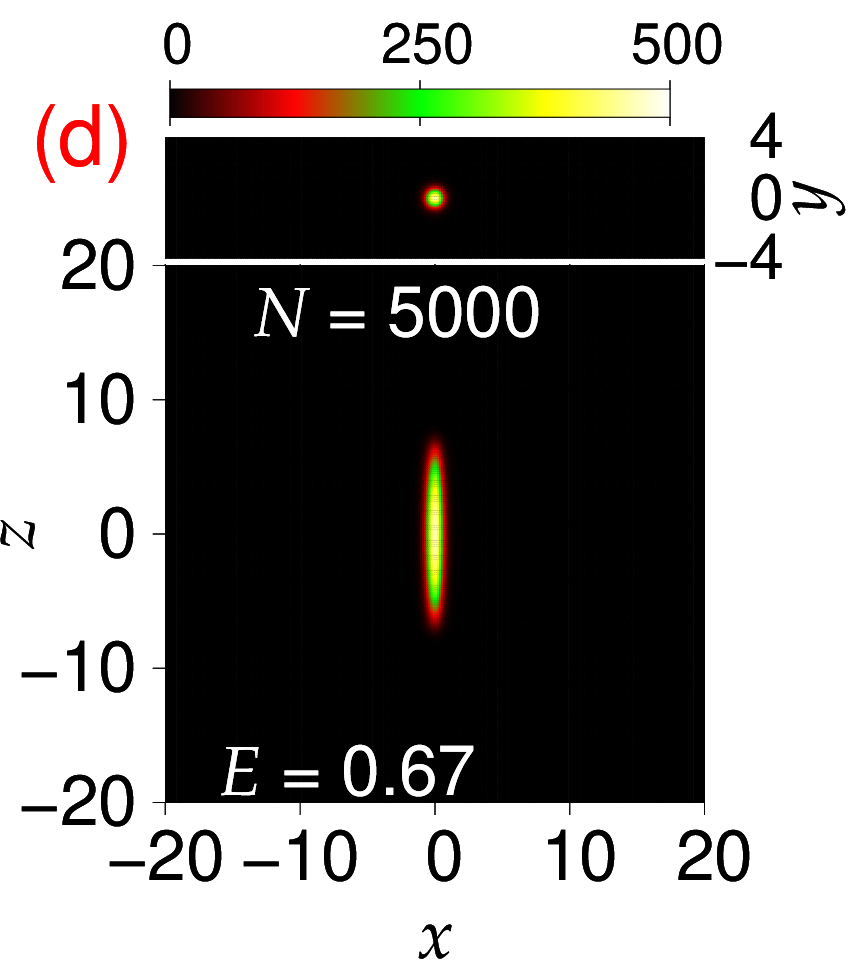}
\includegraphics[width=.32\linewidth]{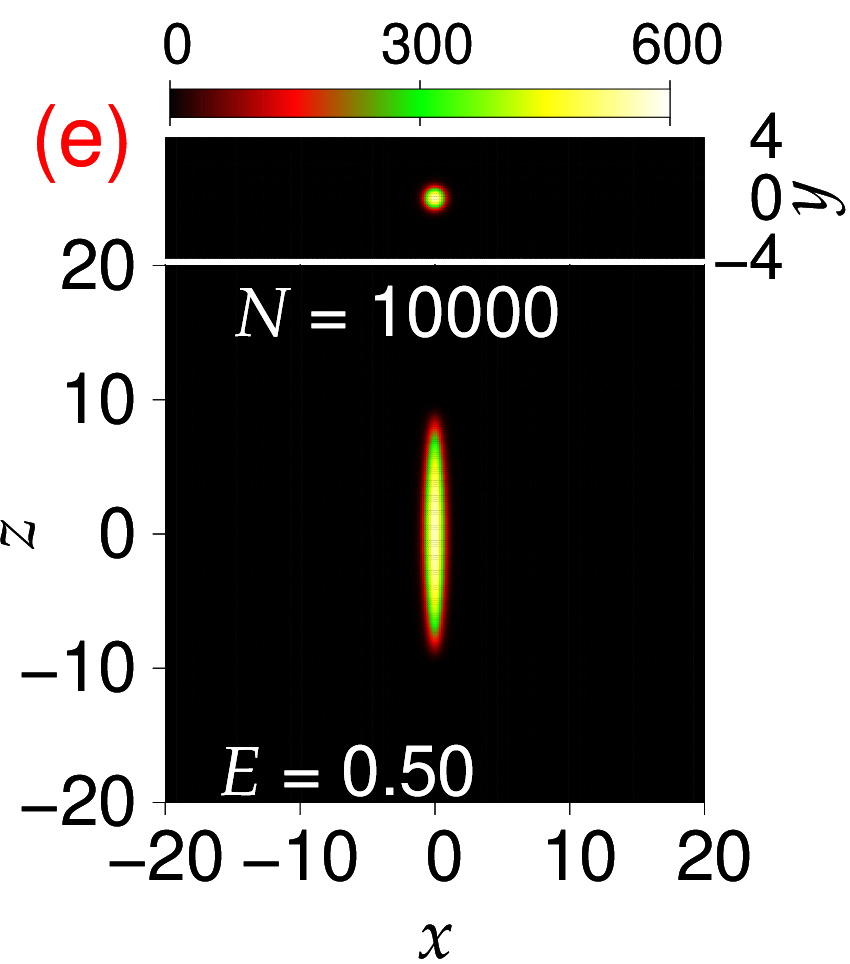}
\includegraphics[width=.32\linewidth]{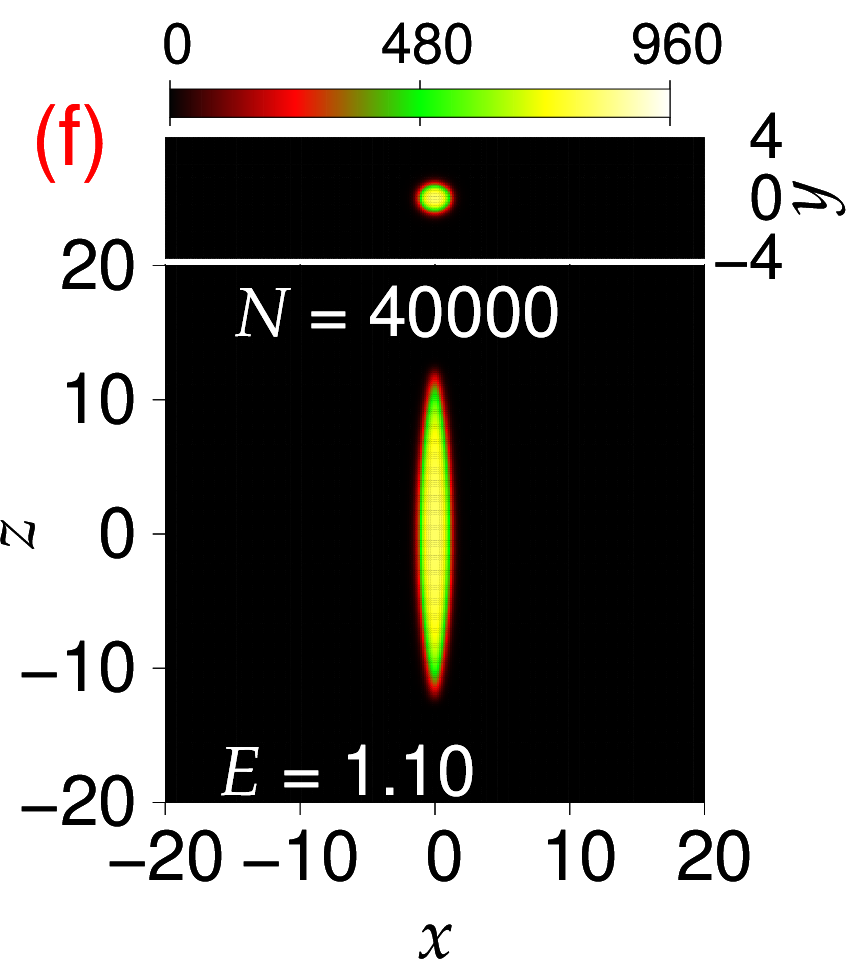} 

\caption{ Contour plot of density $N|\psi(x,y,0)|^2$ (upper panel) and $N|\psi(x,0,z)|^2$ (lower panel)
of a dipolar droplet of (a) $N=500$,  (b) $N=1000$, (c) $N=2000$, (d) $N=5000$, (e) $N=10000$, and (f) $N=40000$    $^{164}$Dy atoms.   The energy per atom  $E$ and the number of atoms $N$ are displayed in the respective plots.
{ Plotted quantities in all figures [except Fig. \ref{fig7})]  are dimensionless}; the length is  expressed in units of $l\equiv\sqrt{\hbar/m\omega_z}=0.6075$ $\mu$m and density in units of $l^{-3}$. Other   parameters  are  $f_x=33/167$, $f_y=1$, and $f_z=60/167$, $a=85a_0/l, a_{\mathrm{dd}}=130.8a_0/l$. 
}
\label{fig1} 
\end{center}
\end{figure}

 For the formation of droplets we need a strongly dipolar atom with $\varepsilon_{\mathrm{dd}}>1$  necessarily \cite{2d3}. In this study, as in Ref. \cite{luis1}, we take $a=85a_0$, close to its experimental estimate $a=(92\pm 8)a_0$  \cite{scatmes},   
and $a_{\mathrm{dd}}=130.8a_0$, where $a_0$  is the Bohr radius; consequently, $\varepsilon_{\mathrm{dd}}=1.5388... >1$.
This value of scattering length is close to the scattering lengths   $a=88a_0$ \cite{2d2exFer,2d4} and $a=70a_0$ \cite{blakieprl}
used in some other  studies of quantum droplets in a quasi-2D dipolar BEC. 
In this study we {take}  $m(^{164}$Dy) $=164 \times 1.66054\times 10^{-27}$ kg, $\hbar =  1.0545718 \times  10^{-34}$ $m^2$ kg/s, $\omega_y = 2\pi \times 167 $  Hz, consequently, the unit of length $l=\sqrt{\hbar/m\omega_y}= 0.6075$  $\mu$m.

 \begin{figure}[t!]
\begin{center}
\includegraphics[width=.32\linewidth]{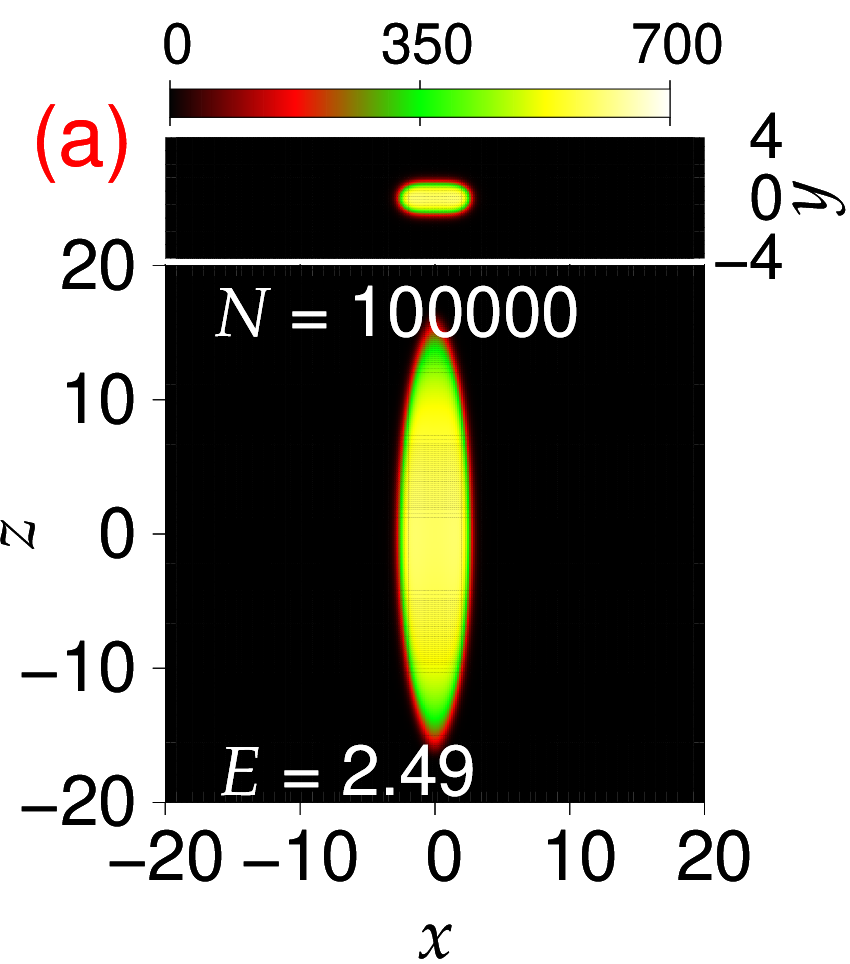}
\includegraphics[width=.32\linewidth]{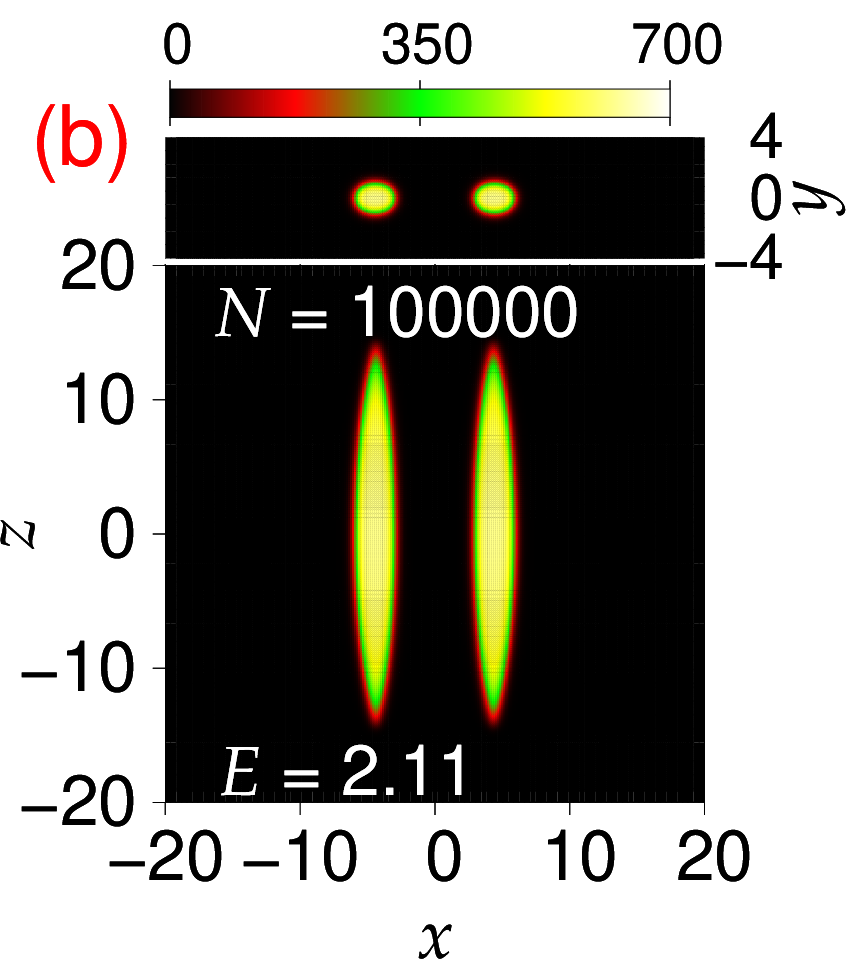}
\includegraphics[width=.32\linewidth]{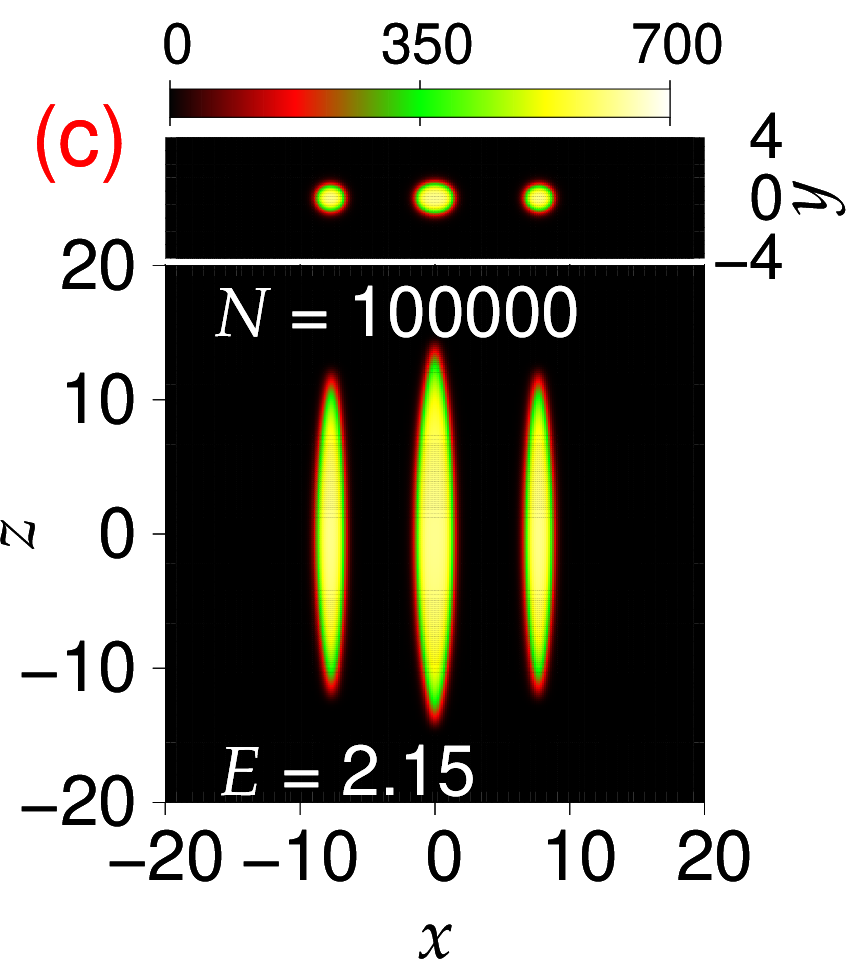}
\includegraphics[width=.32\linewidth]{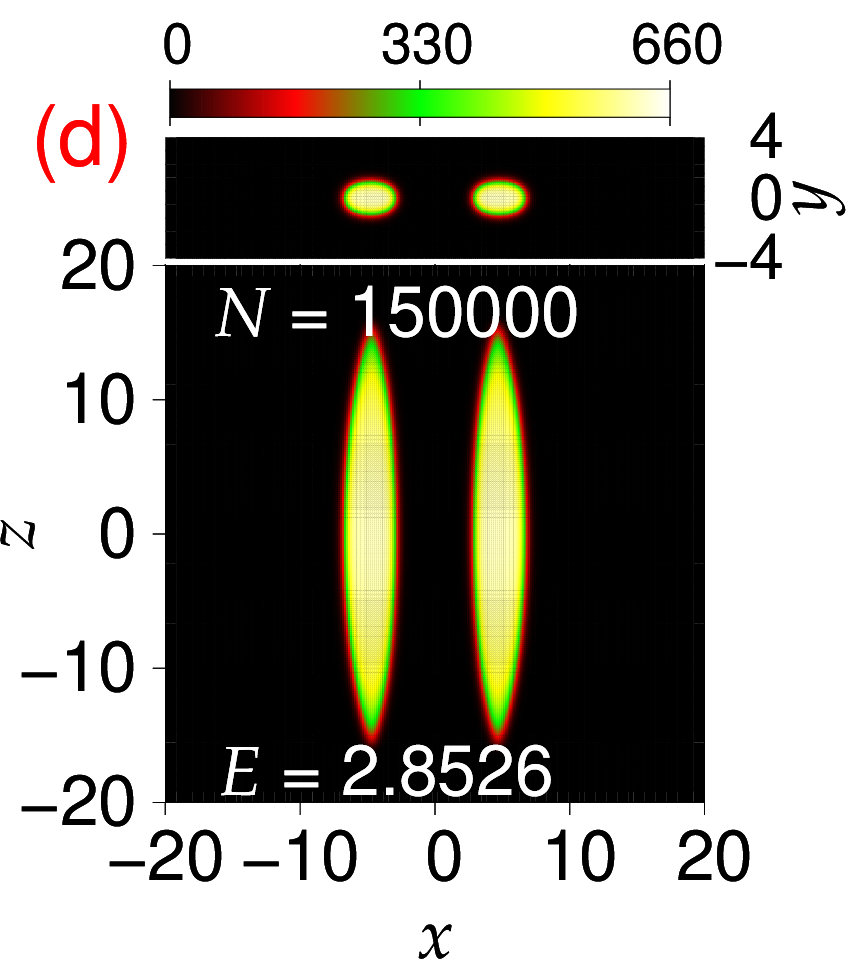}
\includegraphics[width=.32\linewidth]{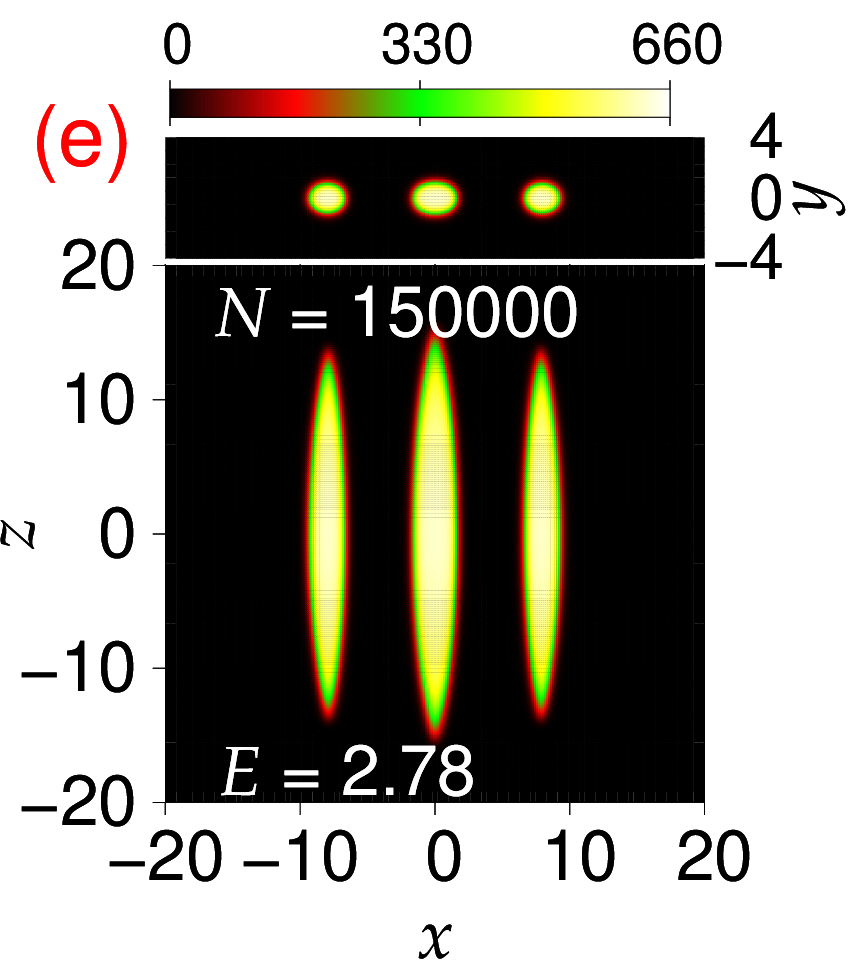}
\includegraphics[width=.32\linewidth]{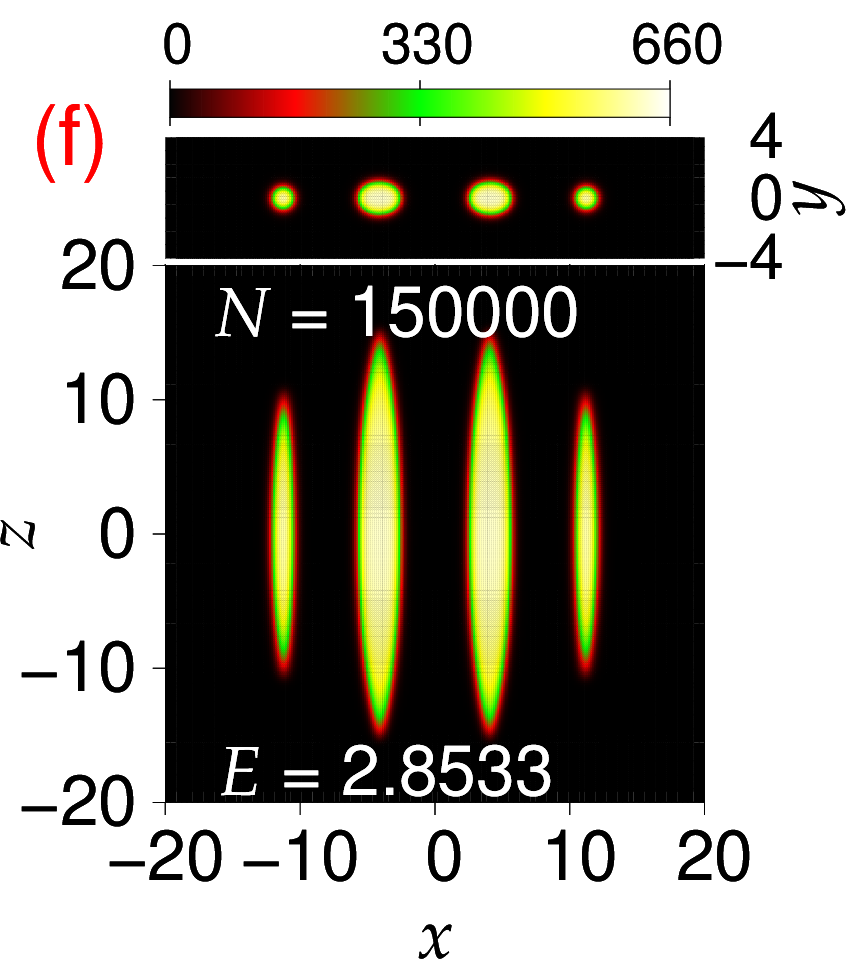}
\includegraphics[width=.32\linewidth]{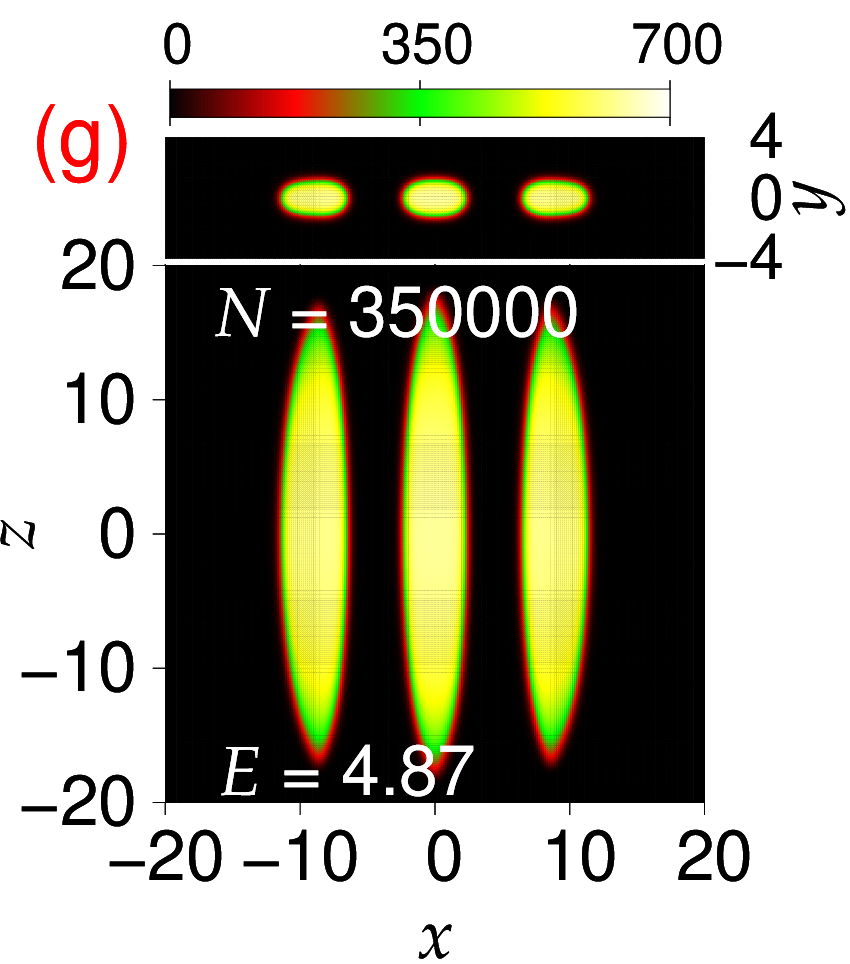}
\includegraphics[width=.32\linewidth]{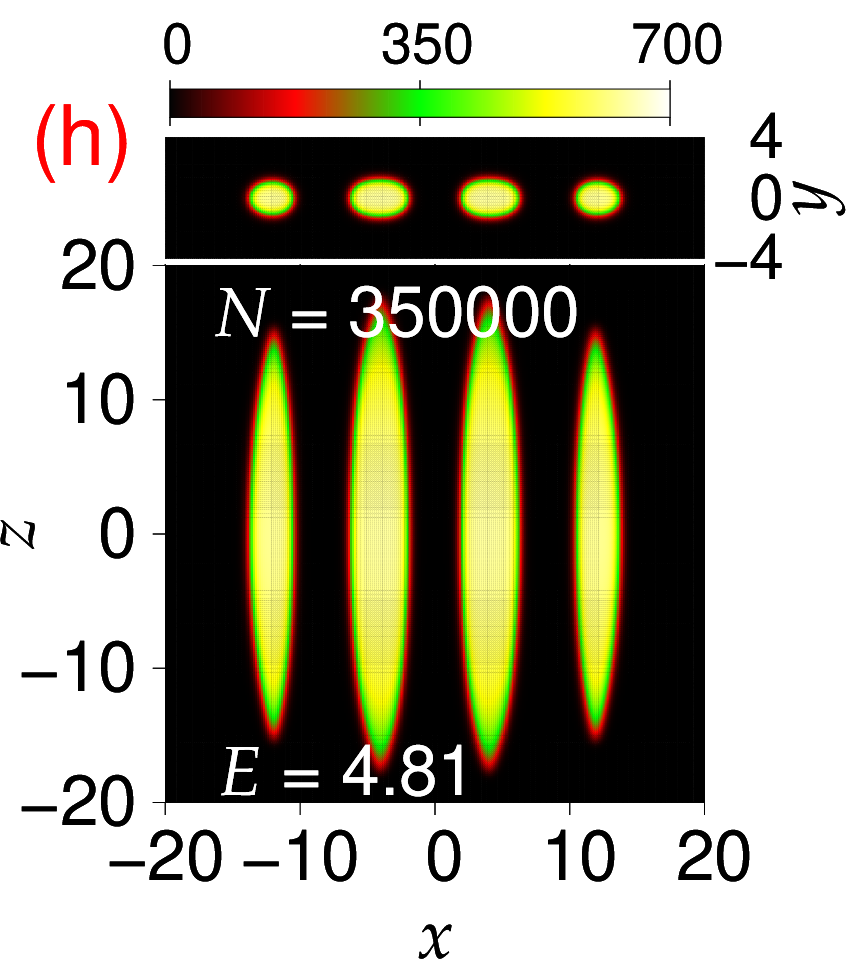}
\includegraphics[width=.32\linewidth]{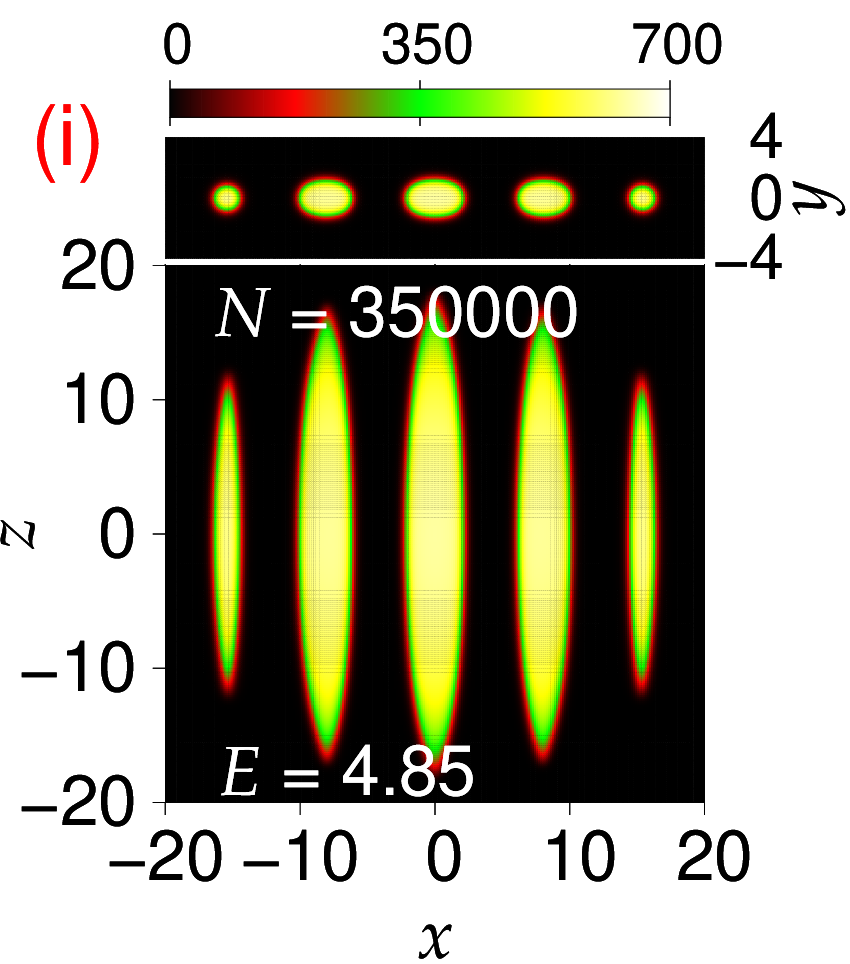}
\includegraphics[width=.32\linewidth]{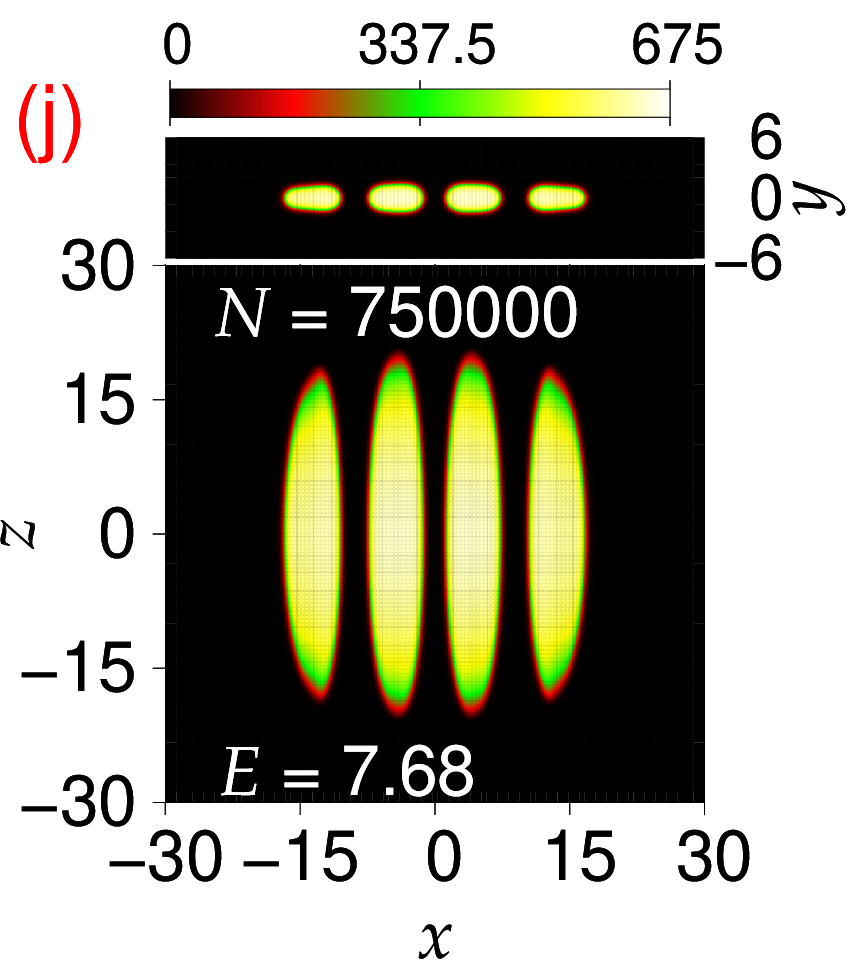}
\includegraphics[width=.32\linewidth]{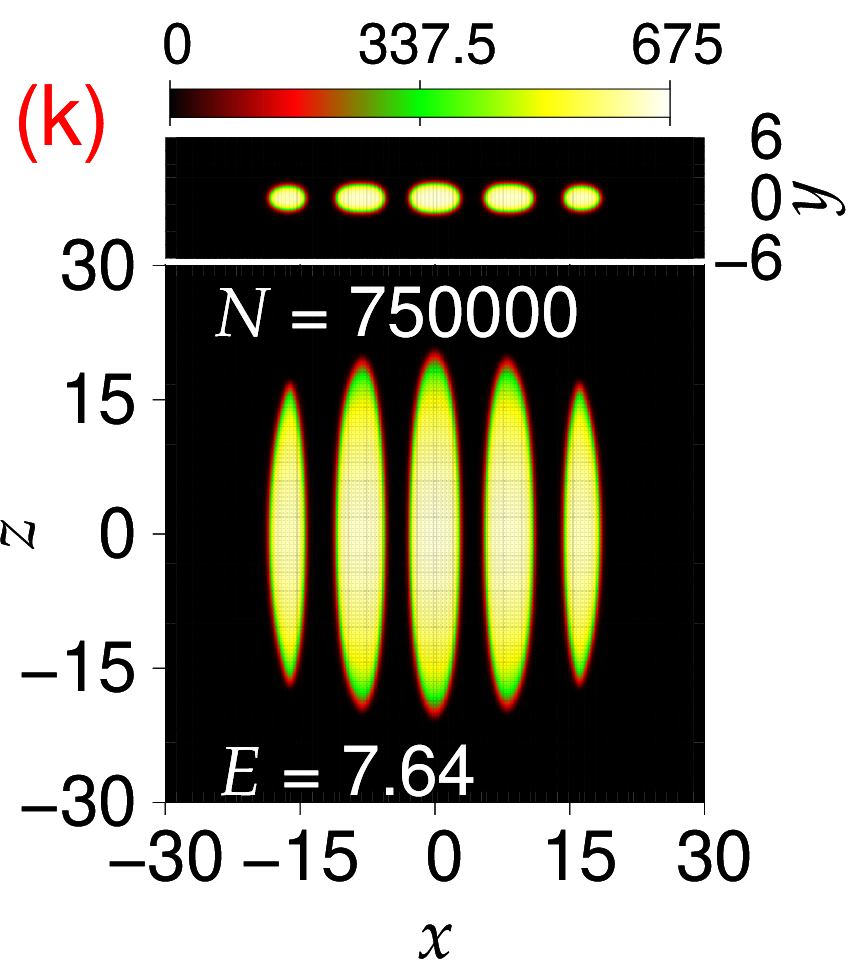}
\includegraphics[width=.32\linewidth]{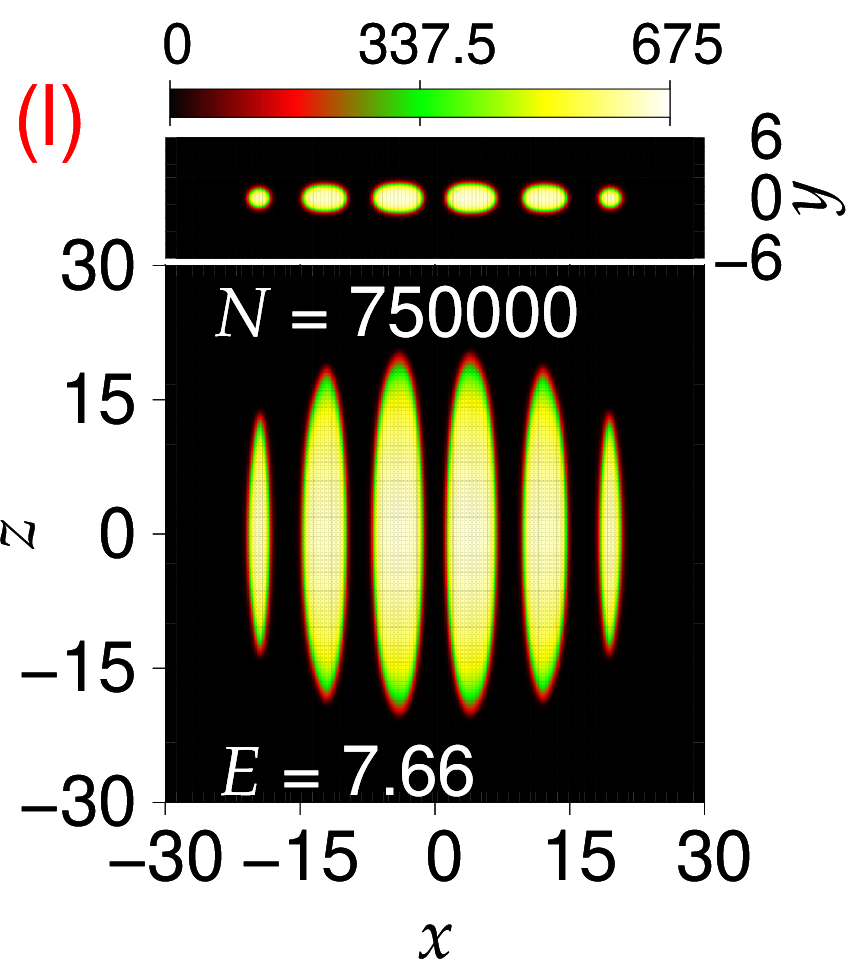}
\includegraphics[width=.32\linewidth]{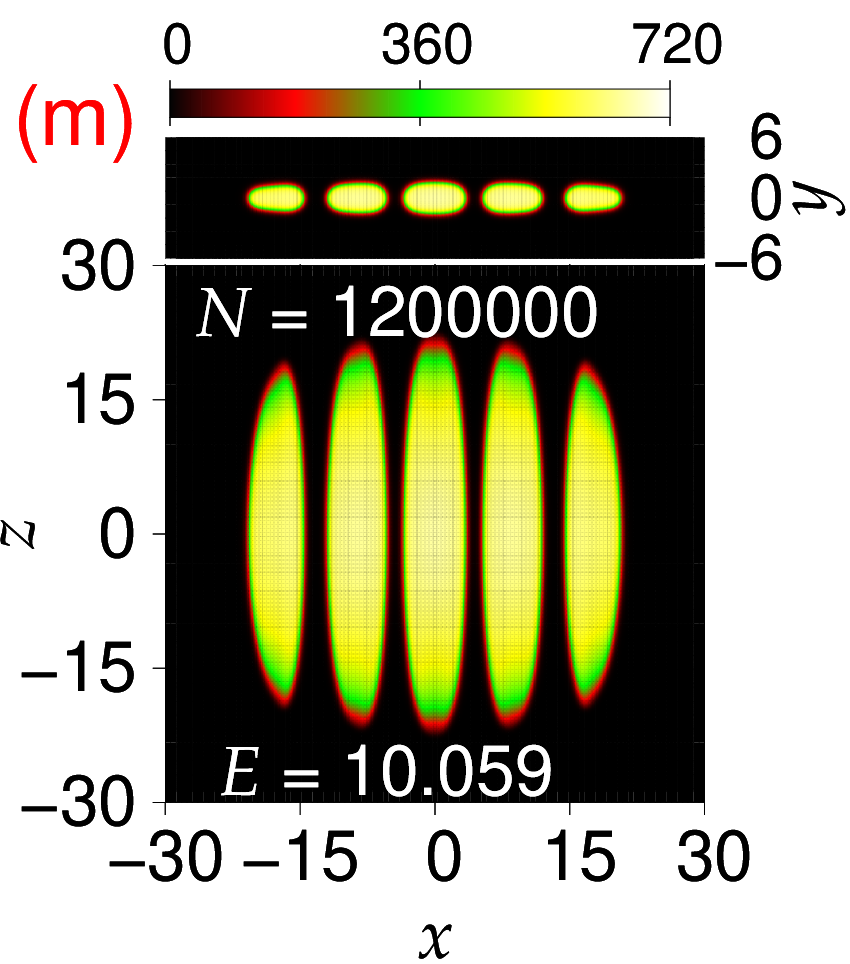}
\includegraphics[width=.32\linewidth]{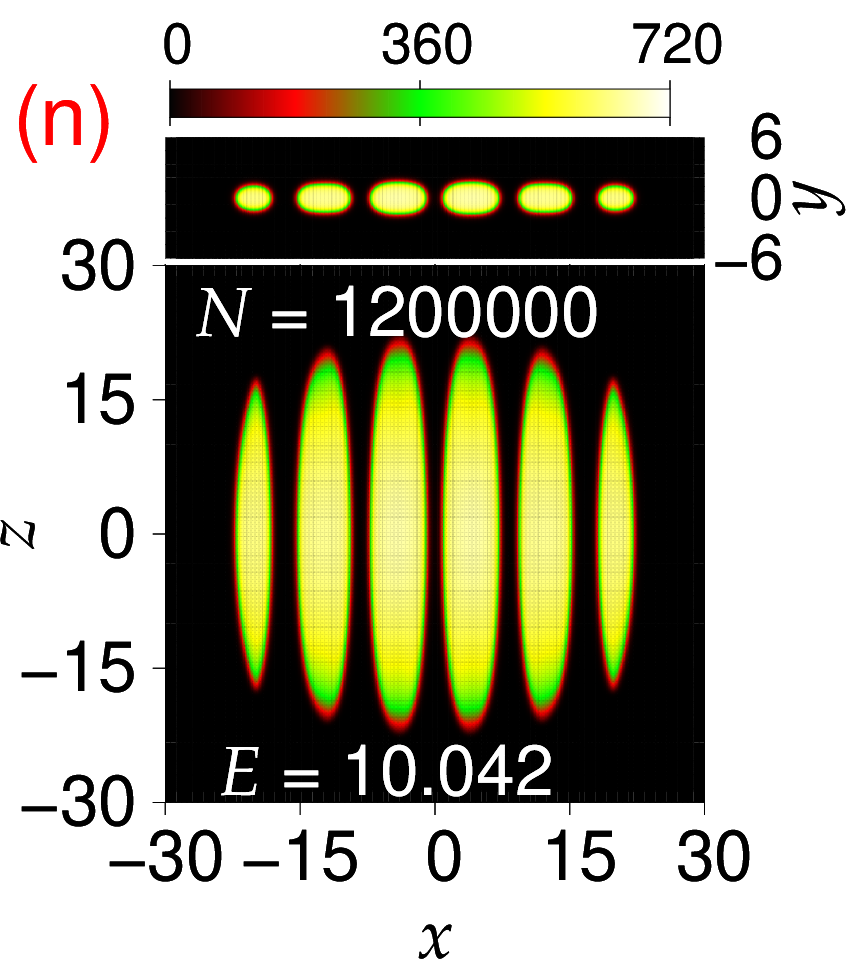}
\includegraphics[width=.32\linewidth]{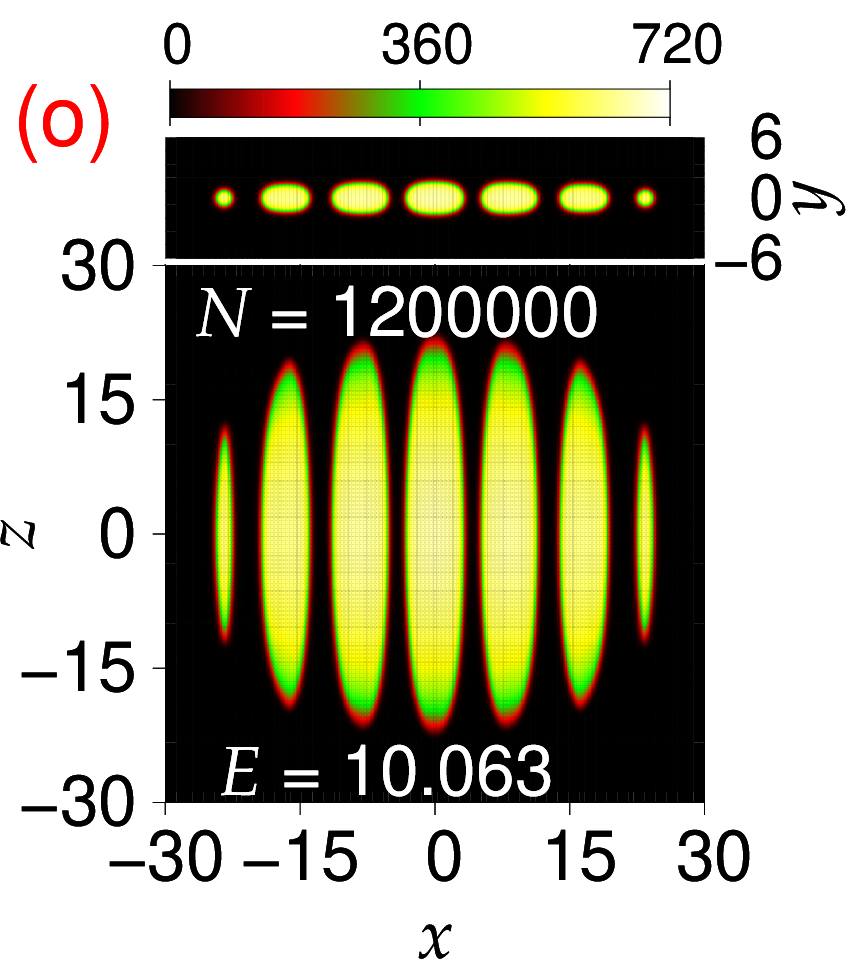} 

\caption{ Contour plot of density $N|\psi(x,y,0)|^2$ (upper panel) and $N|\psi(x,0,z)|^2$
(lower panel) of a    (a)  one-droplet metastable state, (b)  two-droplet ground state and (c) a three-droplet metastable state 
of $N=10^5 $  $^{164}$Dy atoms.  The same of a    (d)  two-droplet metastable state, (e)  three-droplet ground state and (f) a four-droplet metastable state 
of $N=1.5\times 10^5 $  $^{164}$Dy atoms. 
The same of a    (g)  three-droplet metastable state, (h)  four-droplet ground state and (i) a five-droplet metastable state 
of $N=350000 $  $^{164}$Dy atoms. 
The same of a    (j)  four-droplet metastable state, (k)  five-droplet ground state and (l) a six-droplet metastable state 
of $N=750000 $  $^{164}$Dy atoms.
The same of a    (m)  five-droplet metastable state, (n)  six-droplet ground state and (o) a seven-droplet metastable state 
of $N=1200000 $  $^{164}$Dy atoms. All plots are labeled by  respective  $E$ and $N$ values.
Other parameters are  $f_x=33/167$, $f_y=1, f_z = 60/167 $, $a=85a_0/l, a_{\mathrm{dd}}=130.8a_0/l$.
 }
\label{fig2} 
\end{center}
\end{figure}

\begin{figure}[t!]
\begin{center}
\includegraphics[width=.49\linewidth]{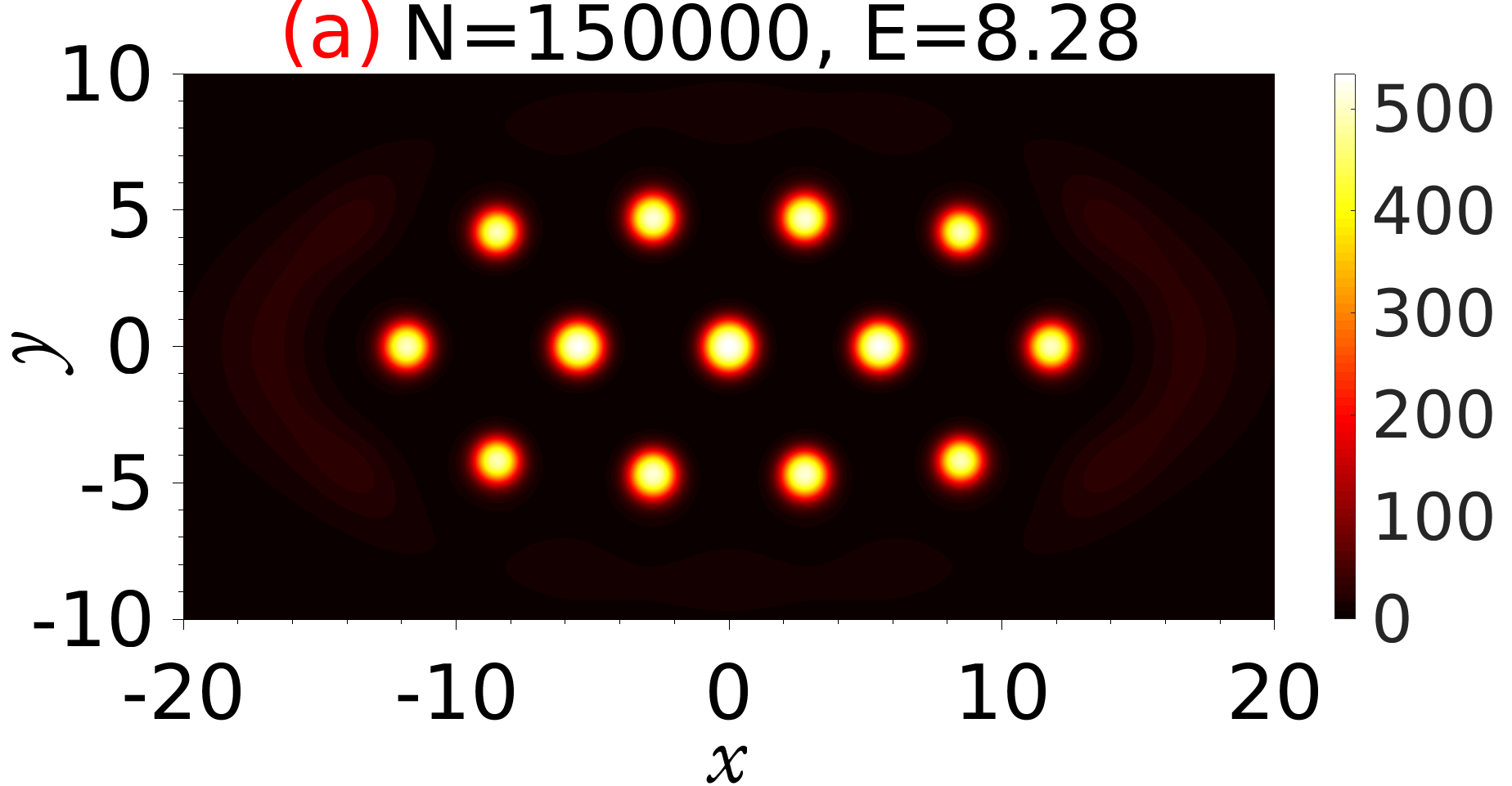}
\includegraphics[width=.49\linewidth]{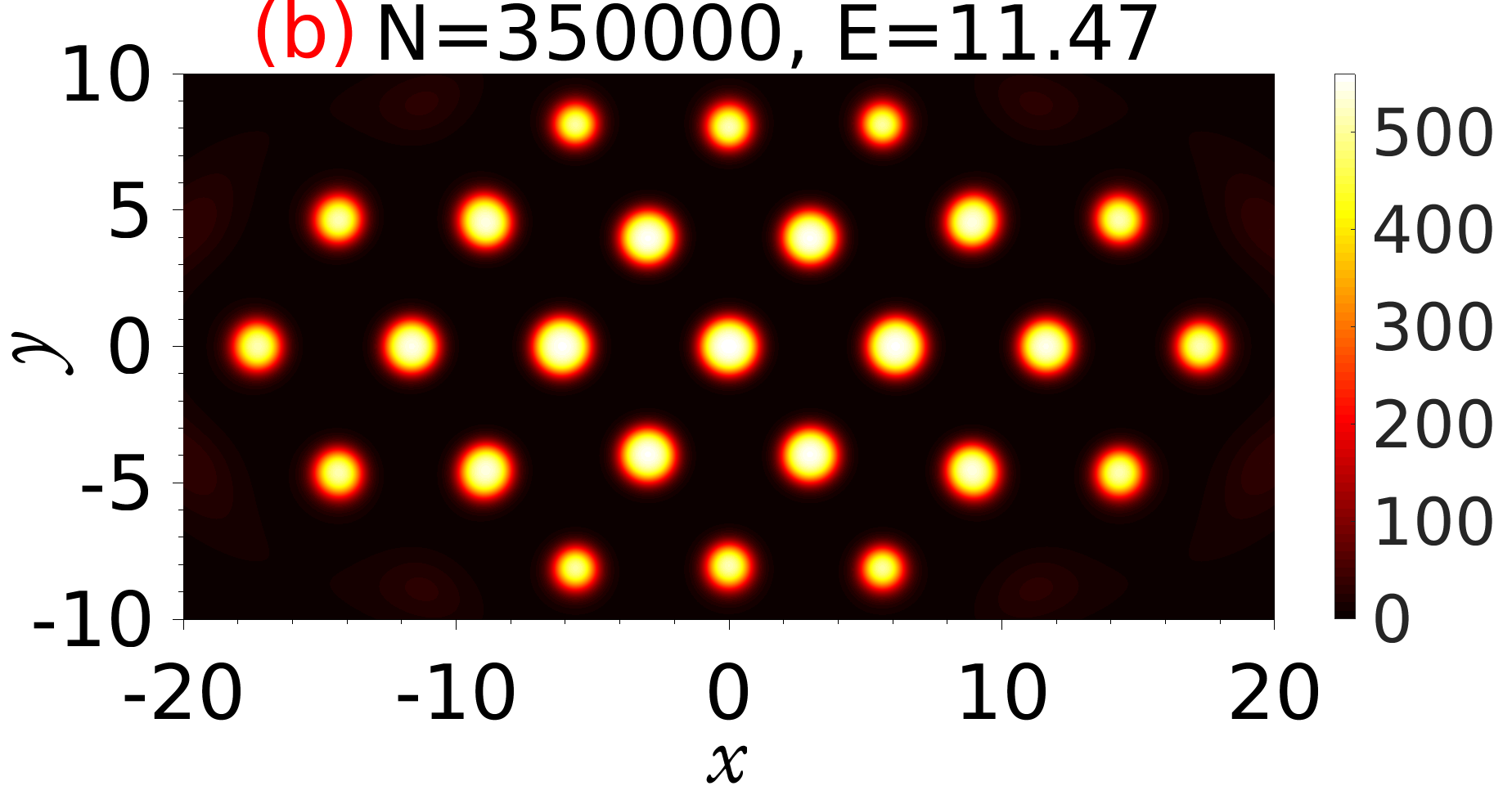}
\includegraphics[width=.49\linewidth]{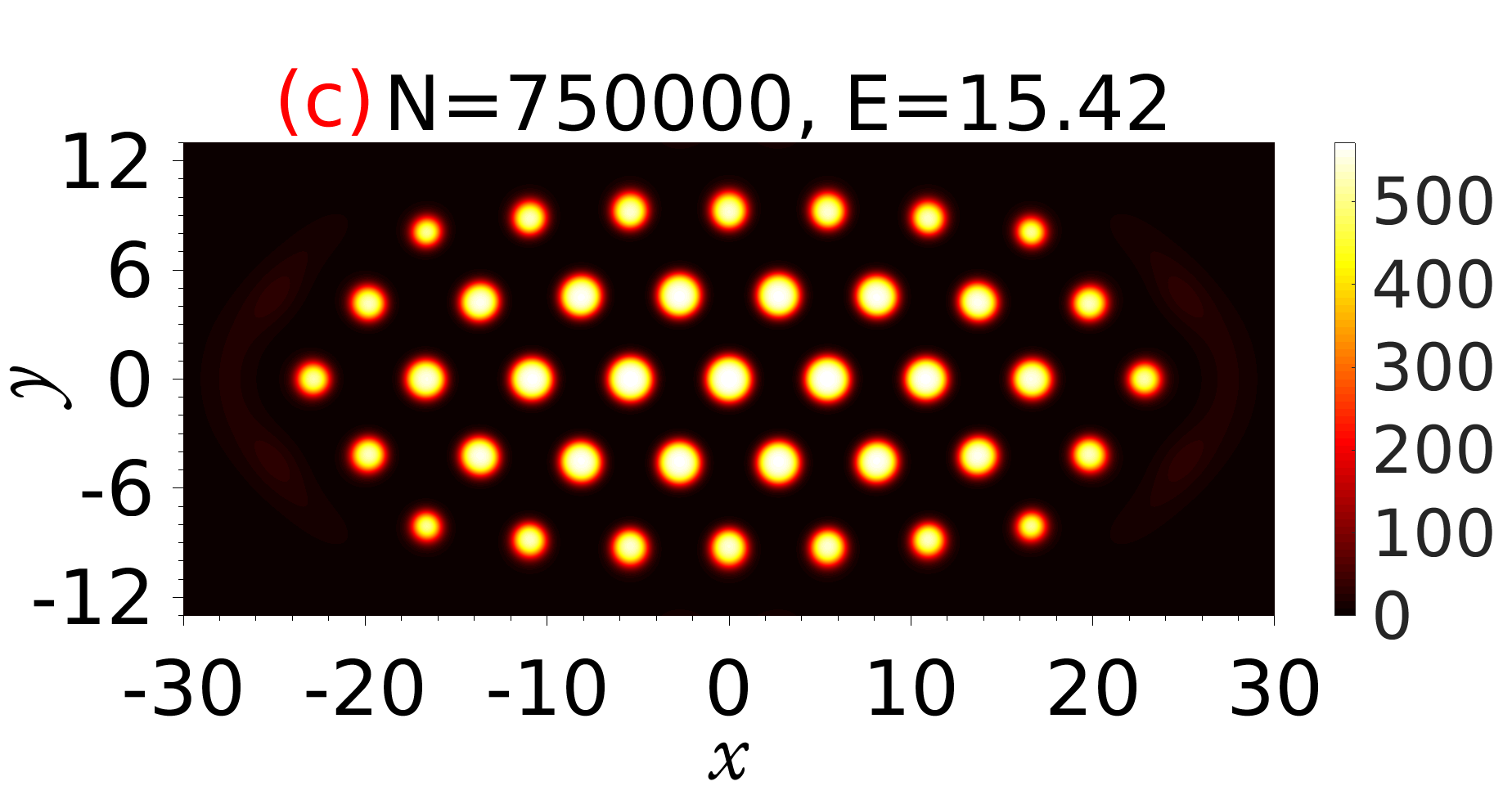}
\includegraphics[width=.49\linewidth]{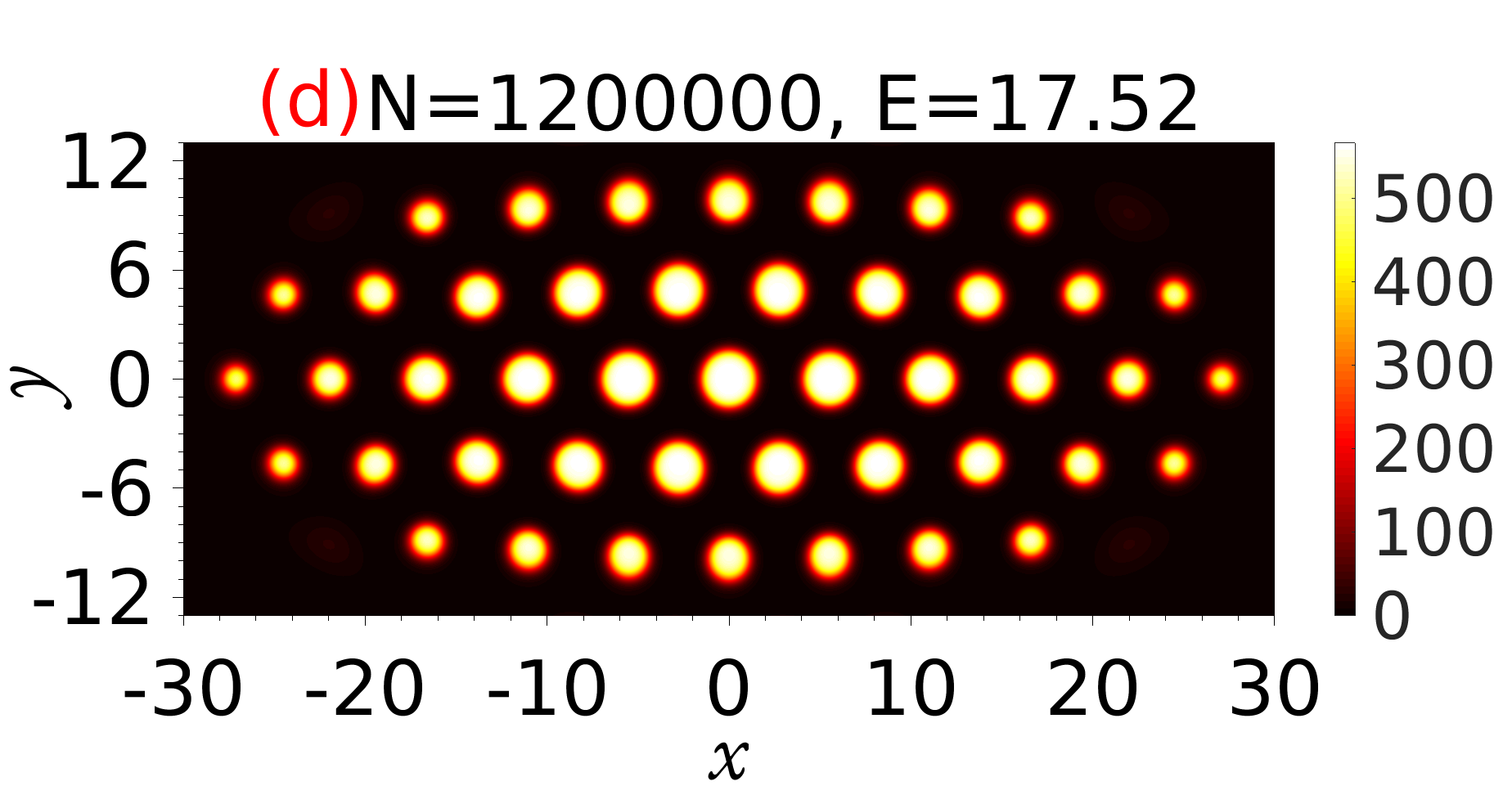}

\caption{ Contour plot of density $N|\psi(x,y,0)|^2$  labeled by energy $E$ and number of atoms $N$
of a  dipolar BEC of   (a)  $N=150000$, (b)   $N=350000$, (c)  $N=750000$, (d) $N=1200000$, $^{164}$Dy  atoms forming a triangular-lattice supersolid.
  Other parameters are  $f_x=33/167$, $f_y=60/167, f_z = 1  
$, $a=85a_0/l, a_{\mathrm{dd}}=130.8a_0/l$.
 }
\label{fig3} 
\end{center}
\end{figure}

First we study the formation of a single droplet. For a very small number ($N \lessapprox 500$)  of atoms we have a normal BEC over an extended region.  In Fig. \ref{fig1} we display a contour plot of densities $N|\psi(x,y,0)|^2$ (top panel) and $N|\psi(x,0,z)|^2$ (bottom panel) for (a) $N=500$, (b) $N=1000$, (c) $N=2000$, (d) $N=5000$, (e) $N=10000$, and (f) $N=40000$. 
For $N=500$ the BEC is more like in a transition regime from a normal BEC to a droplet with shrinking size as illustrated in Fig. \ref{fig1}(a).
A single droplet can be formed for a small number ($N=1000$) of  atoms, viz. Fig. \ref{fig1}(b), in the  present trap $f = (33/167, 1,60/167)$ Hz, which we call a mini droplet.  The droplet has very small size in the $x$-$y$ plane and is elongated along the polarization $z$ direction in the quasi-2D $x$-$z$ plane. This quasi-2D droplet  is of different nature from the one studied previously 
\cite{2d2exFer,2d4,luis1}: the difference between the two cases is the swapped frequencies $f_y  \leftrightarrow  f_z$.  With all other parameters unchanged, the minimum number of atoms for forming a single droplet in these previous studies 
is $N\approx 20000$, with trap frequencies $f = (33, 60,167)$ Hz, viz. Fig. \ref{fig1}(b) of Ref. \cite{luis1},
compared to the present number $N\approx 1000$. Nevertheless, in both cases the droplet is elongated in the $z$ direction with a shrinked shape in the $x$-$y$ plane, viz. Fig. \ref{fig1}.   The droplet becomes thicker and longer as the number of atoms $N$ increases as can be seen in Figs. \ref{fig1}(c)-(f) for $N=2000, 5000,10000,40000$. Up to about $N=50000$ the only possible state found in imaginary-time propagation is the single-droplet state of Fig. \ref{fig1}. Beyond this, for $50000 \lessapprox N \lessapprox 90000$ a metastable two-droplet state is also possible  
in addition to the one-droplet ground state.

As $N$ is further increased beyond $N\gtrapprox 90000$, the two-droplet state becomes the stable ground state and the one-droplet state becomes a metastable excited state.  In addition, a three-droplet metastable  state appears as illustrated in Fig. \ref{fig2}(a)-(c) for $N=100000$ through a contour plot of densities
$N|\psi(x,y,0)|^2$ (upper panel) and $N|\psi(x,0,z)|^2$ (lower panel). The quasi-1D array of droplets is clearly visible  in the $x$-$y$ plane.  Although, the trap is of quasi-2D type ($f_y \gg f_x,f_z$) in the $x$-$z$ plane, the realized supersolid is of quasi-1D type, e.g. an array of quasi-1D droplets along the $x$ direction.  As $N$ is further increased more droplets appear.
For $N=150000$, there are three states:  a three-droplet ground state with a two-droplet  and a four-droplet metastable state as can be found in Figs. \ref{fig2}(d)-(f).  
For $N=350000$, the three possible states are the four-droplet 
ground state and a  three-droplet  and a five-droplet metastable state, viz. Figs. \ref{fig2}(g)-(i). 
For $N=750000$ the states are the five-droplet ground state and a four-droplet and a six-droplet metastable state as shown in Figs. \ref{fig2}(j)-(l). 
For $N=1200000$ we have the six-droplet ground state and a five- and a seven-droplet metastable state as depicted in Figs. \ref{fig2}(m)-(o).  The droplets of Fig. 2 can accommodate a very large number of atoms $-$ between $50000$ and $200000$. We call these droplets mega droplets.  In comparison, the previously studied normal droplets  \cite{2d2exFer,2d4,luis1}, could accommodate a small number of atoms $-$ between $N=5000$ and $N=15000$.  In the present scenario, for a fixed $N$, usually there are one ground state and one or two metastable states. In previous studies for a fixed $N$, there could be many close-by metastable excited states \cite{luis1}. In the quasi-2D $x$-$z$ plane (with a strong trap in the $y$ direction)  the droplets form a prominent stripe pattern, whereas in previous studies \cite{2d4,luis1} in the quasi-2D $x$-$y$ plane (with a strong trap in the polarization $z$ direction) the droplets are arranged in a quasi-2D triangular, square or other form of lattice.

\begin{figure}[t!]
\begin{center}
\includegraphics[width=.4\linewidth]{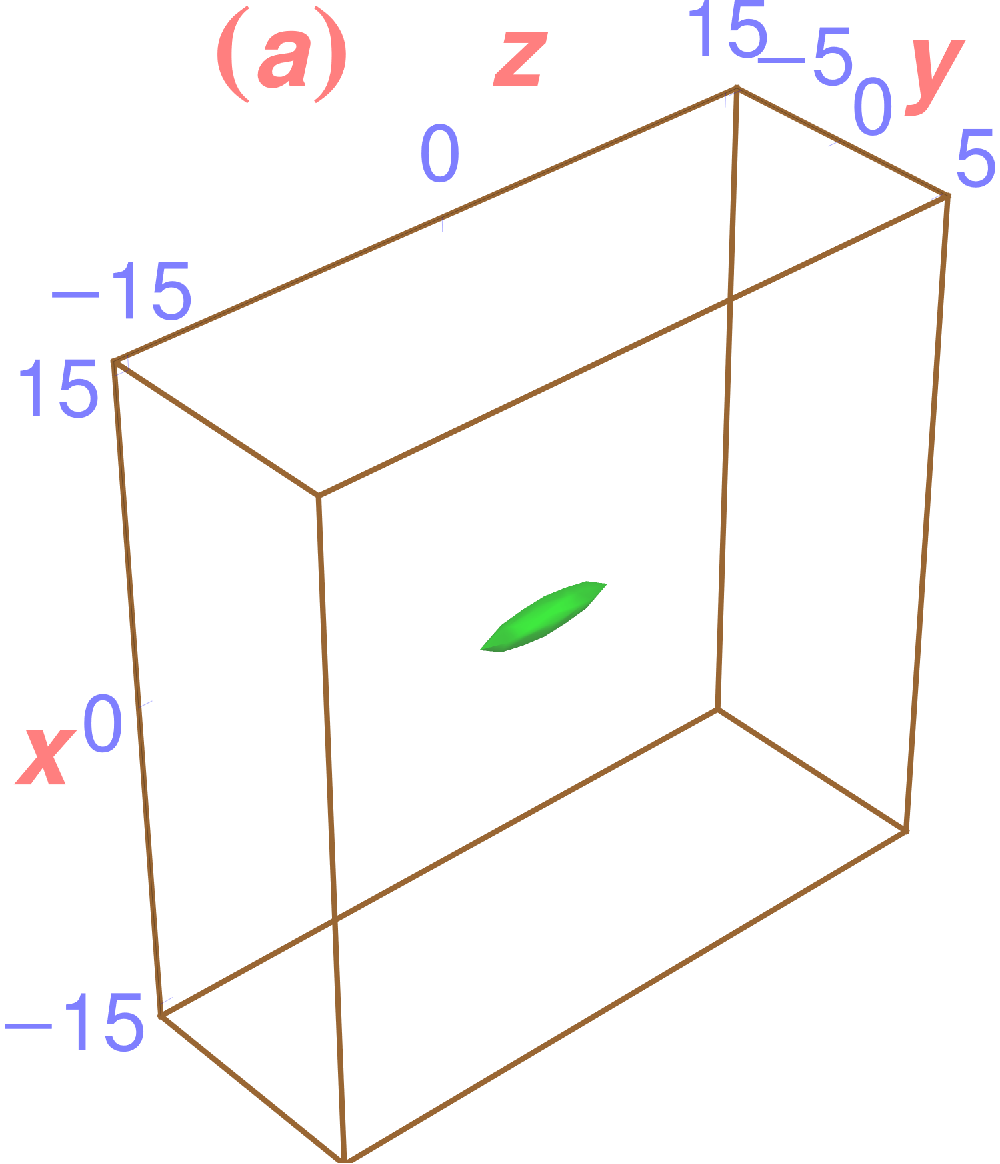}
\includegraphics[width=.4\linewidth]{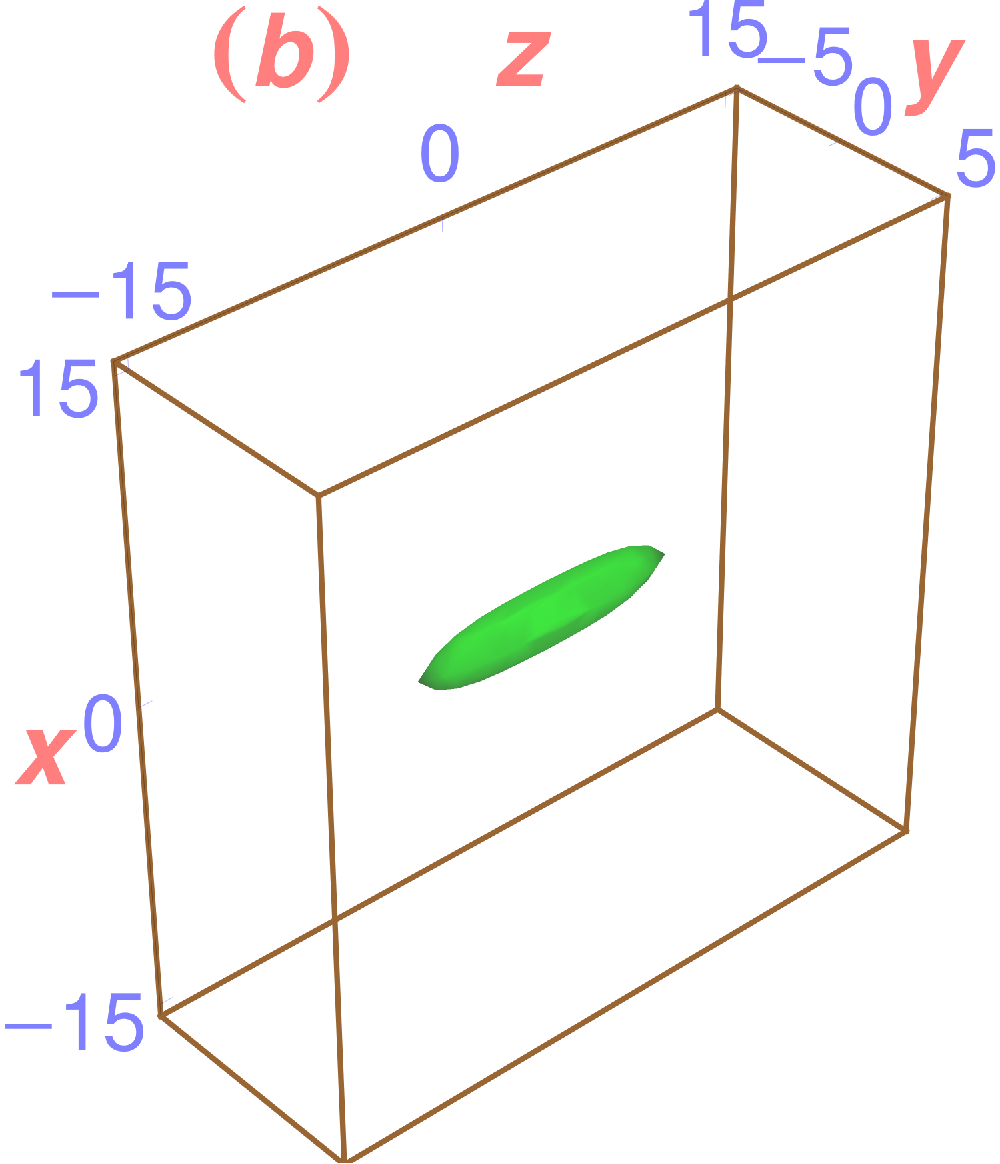}
\includegraphics[width=.4\linewidth]{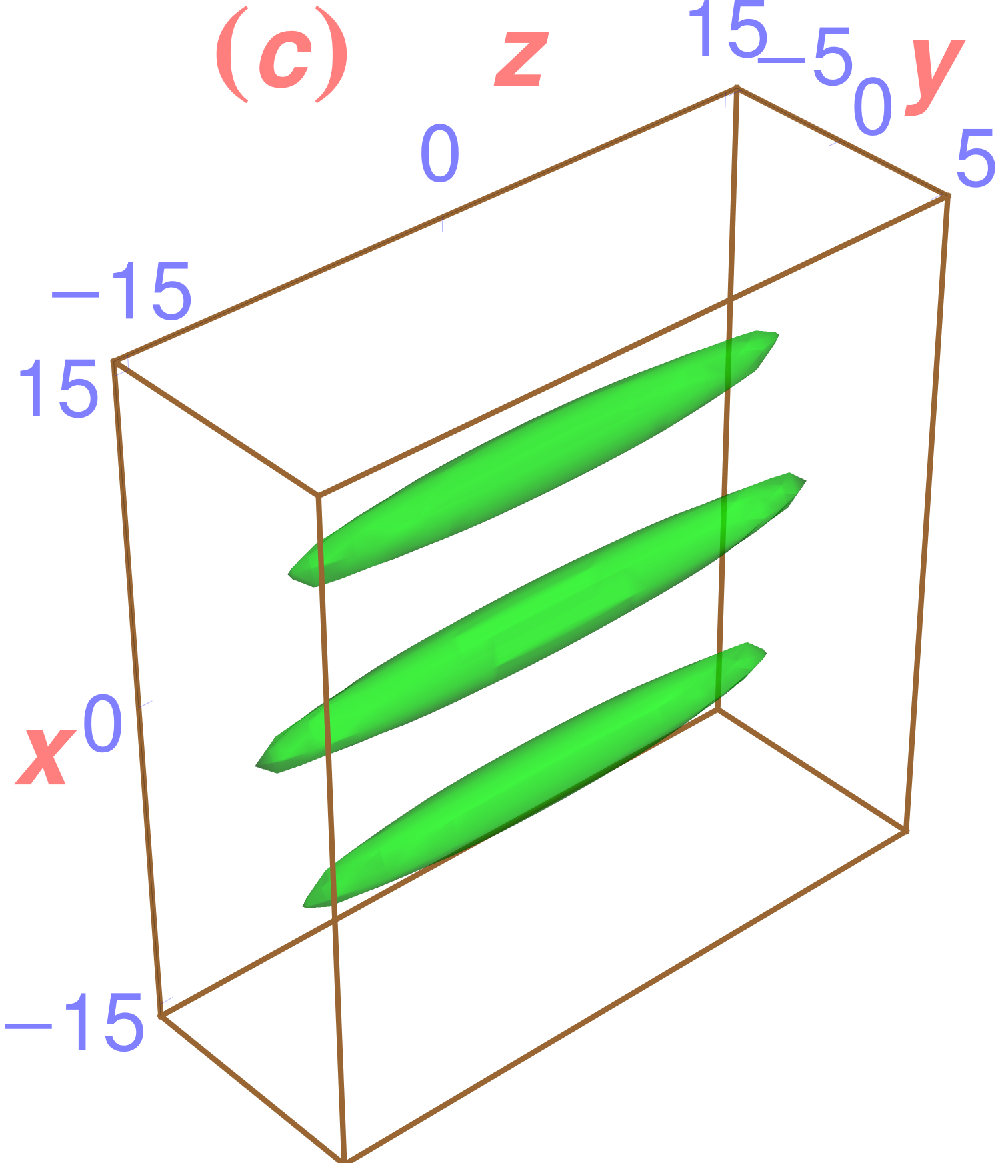}
\includegraphics[width=.4\linewidth]{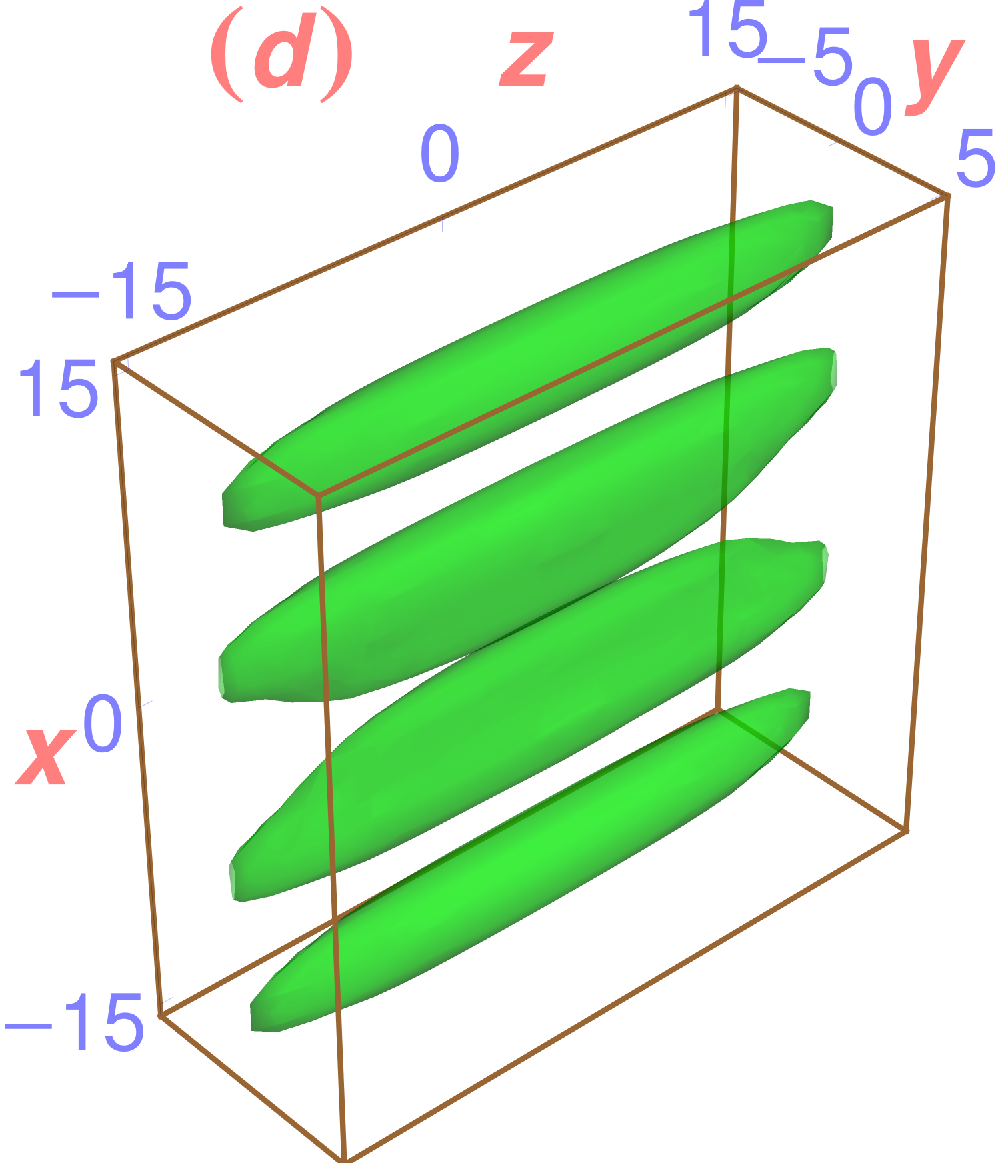}

\caption{  Three-dimensional isodensity plot of  $N|\psi(x,y,z)|^2$  
of a  dipolar BEC of   (a)  $N=2000$, (b)   $N=5000$, (c)  $N=150000$, (d) $N=350000$, $^{164}$Dy  atoms.
  Other parameters are  $f_x=33/167$, $f_y=1, f_z = 60/167  
$, $a=85a_0/l, a_{\mathrm{dd}}=130.8a_0/l$. The value of density on contour is 200.
 }
\label{fig4} 
\end{center}
\end{figure}

In Fig. \ref{fig3} we present the triangular-lattice supersolid state, in the case with the $y$ and $z$ traps interchanged with respect to the present case in Fig. 
 \ref{fig2}, through a contour plot of density $N|\psi(x,y,0)|^2$  for (a) $N=150000,$ (b) $N=350000$, (c) $N=750000$, and (d) $N=1200000$. The trap frequencies in the case  of Fig. \ref{fig3} are $f_x= 33/167$,   $f_y=60/167$, and $f_z=1$, whereas those in Fig. \ref{fig2} are  $f_x= 33/167$,   $f_z=60/167$, and $f_y=1$ with the same overall trapping $\bar f =\sqrt[3] {f_xf_yf_z}$.  If we compare Figs. \ref{fig2}(e) and \ref{fig3}(a),  
  Figs. \ref{fig2}(h) and \ref{fig3}(b),
   Figs. \ref{fig2}(k) and \ref{fig3}(c),
      Figs. \ref{fig2}(n) and \ref{fig3}(d),
we find that, although the overall trap frequency $\bar f$ in the two cases are equal, the number of droplets 
  and their arrangement are completely distinct in the two cases. For example,  from   Figs. \ref{fig2}(k) and \ref{fig3}(c) we find that for $N=750000$ in Fig. \ref{fig2} of this study we have 5 droplets whereas in the 
  $f_y\leftrightarrow f_z$ interchanged trap we have 39 droplets. In the first case the droplets are arranged in a quasi-1D array and in the second case they are arranged in a triangular lattice.

\begin{figure}[t!]
\begin{center}
\includegraphics[width=.49\linewidth]{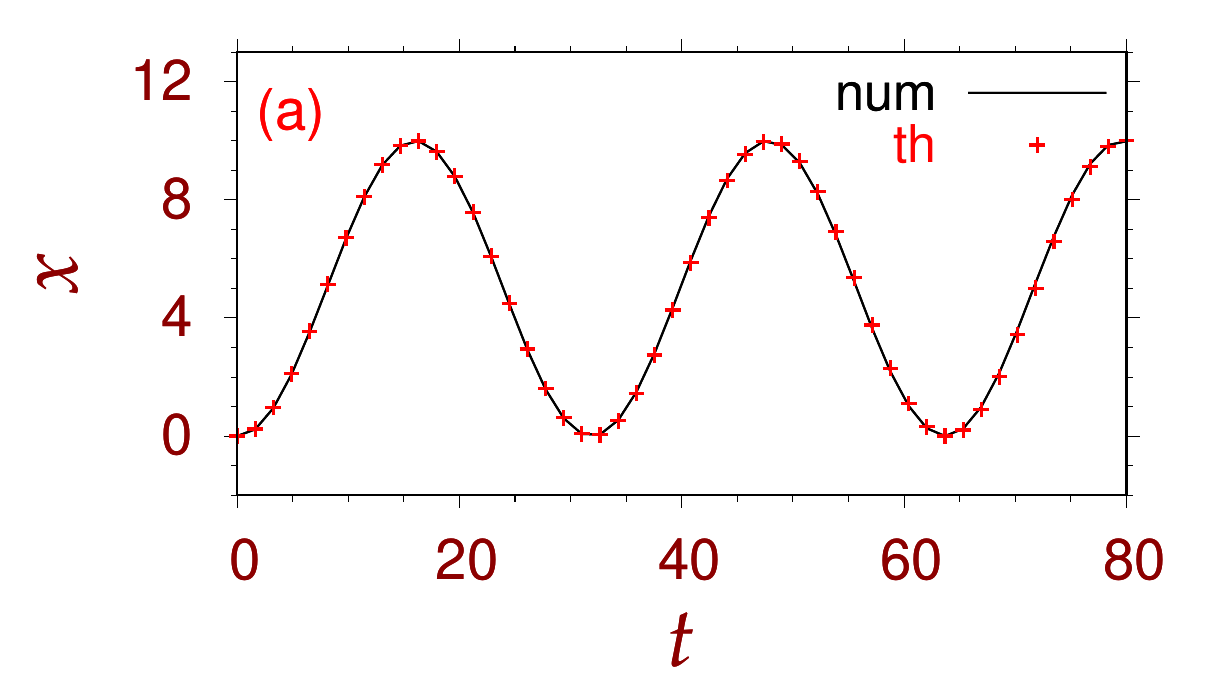}
\includegraphics[width=.49\linewidth]{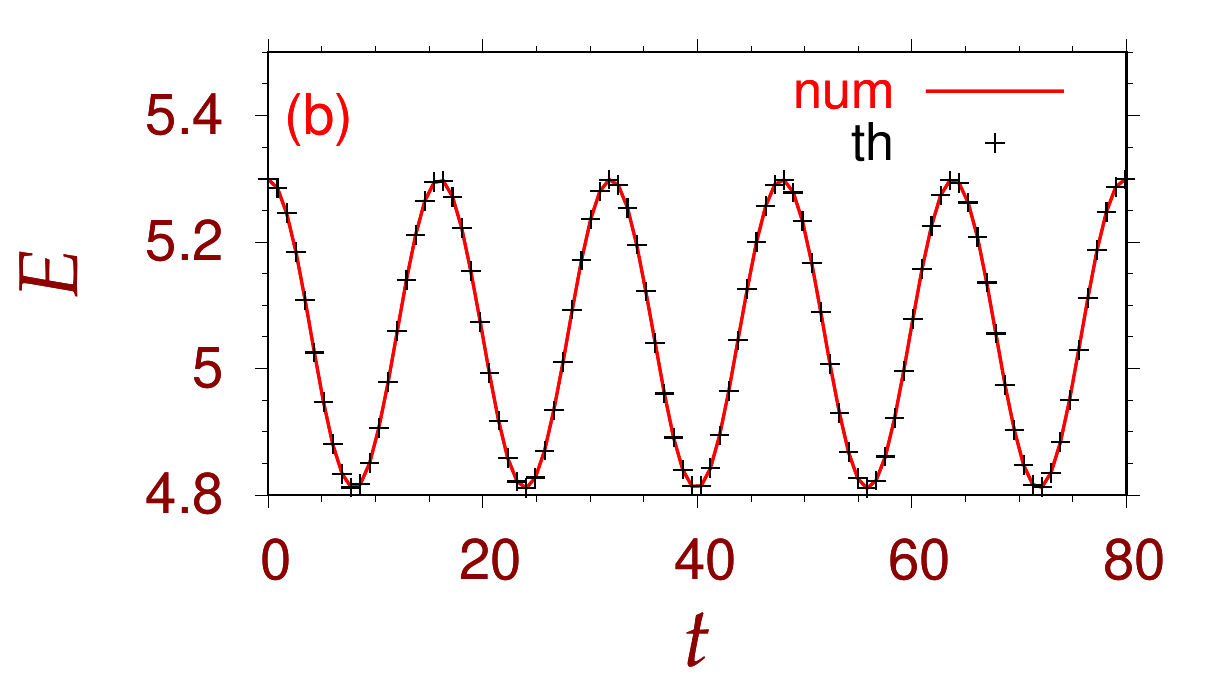}

\caption{(a) Linear displacement $x$ of the four-droplet supersolid  of $N=350000$ $^{164}$Dy atoms  of Fig. \ref{fig2}(h)  versus time $t$ (num)
 executing  {dipole-mode}  oscillation along $x$ direction
as obtained by real-time propagation   fitted to the theoretical prediction  $\cos(f_xt)$ (th).  
 The oscillation is started by  a linear displacement  of $x_0=5$ of  the trap in Eq. (\ref{dip}) at $t=0$.
 (b) Energy $E$ versus time $t$  during this   oscillation 
  fitted to the theoretical prediction  $\cos(2f_xt)$ (th).
  Other parameters are  $f_x=33/167$, $f_y=1, f_z = 60/167 $, $a=85a_0/l, a_{\mathrm{dd}}=130.8a_0/l$.}
\label{fig5} 
\end{center}   
\end{figure}

  To compare the sizes of a mini droplet and a mega droplet we next  consider a 3D isodensity plot of 
 dipolar  single droplets and supersolids for different $N$. In Fig. \ref{fig4} we display such isodensity plot 
  of $N|\psi(x,y,z)|^2$  for (a) $N=2000$, (b) $N=5000$, (c) $N=150000$, and (d) $N=350000$.
Of these (a) and (b) represent a single droplet and (c) and (d) represent quasi-1D dipolar supersolids. With an increase of number of atoms, the size of the droplet increases, the size of a single droplet in Fig. \ref{fig4}(d) being much larger than the single droplet  in   Fig. \ref{fig4}(a). The number of atoms in a droplet of Fig. \ref{fig4}(d), about 80000, is much larger than the same (2000) in  Fig. \ref{fig4}(a). As the total number of atoms is further increased, the number of atoms per droplet increases beyond this number. For example, the average number of atoms per droplet corresponding to the supersolid of Fig. \ref{fig2}(n) is about 200000. 
Thus the droplets of Figs. \ref{fig4}(a)-(b) are termed mini droplets  and those of Figs. \ref{fig4}(c)-(d) termed mega droplets.

  To investigate the dynamical stability of the quasi-1D droplet-lattice state \cite{dyna,related}, 
we now study dipole-mode and scissors-mode oscillations of the present quasi-1D supersolids.  Such oscillation 
tests the rigidity of the supersolid as well as its superfluidity.
 First, we study the dipole-mode oscillation of the quasi-1D four-droplet supersolid of Fig. \ref{fig2}(h) initiated by displacing the trap along the $x$-axis through a distance $x_0=5$.  The initial configuration in this study is the converged stationary-state wave function obtained  by imaginary-time propagation.  The dynamics is studied by real-time propagation in the displaced trap (\ref{dip}). The dipolar supersolid executes a sustained dipole-mode oscillation along the $x$ direction with an amplitude of 5. 
 We display in Fig. \ref{fig5}(a)  
 In Fig. \ref{fig5}(a)  we illustrate  the numerical time evolution of position $x$ of the supersolid
 as well as its theoretical prediction of periodic 
oscillation with the trap frequency $f_x$.
%compare the time evolution of position $x$ of the supersolid  with
% its theoretical prediction of periodic 
%oscillation with the trap frequency $f_x$. 
In Fig. \ref{fig5}(b) we  display a  steady simple-harmonic
oscillation of
the energy of the oscillating supersolid.  The frequency of energy oscillation is double that of the frequency of position oscillation.

\begin{figure}[t!]
\begin{center}
\includegraphics[width=.32\linewidth]{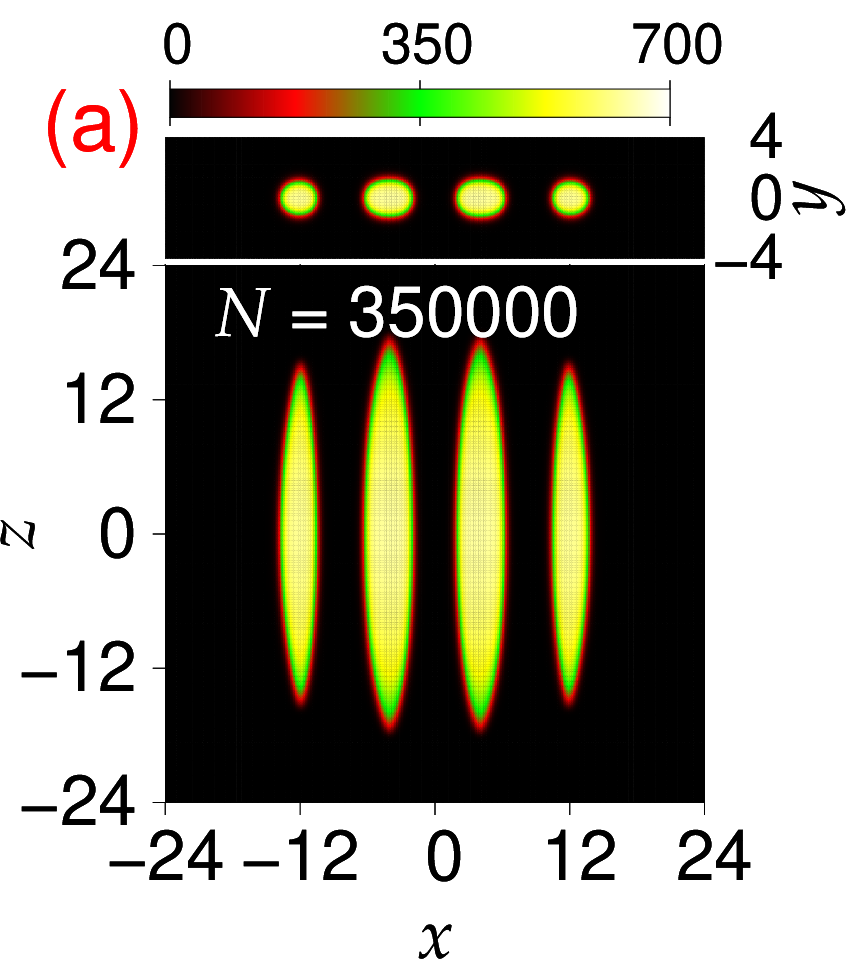}
\includegraphics[width=.32\linewidth]{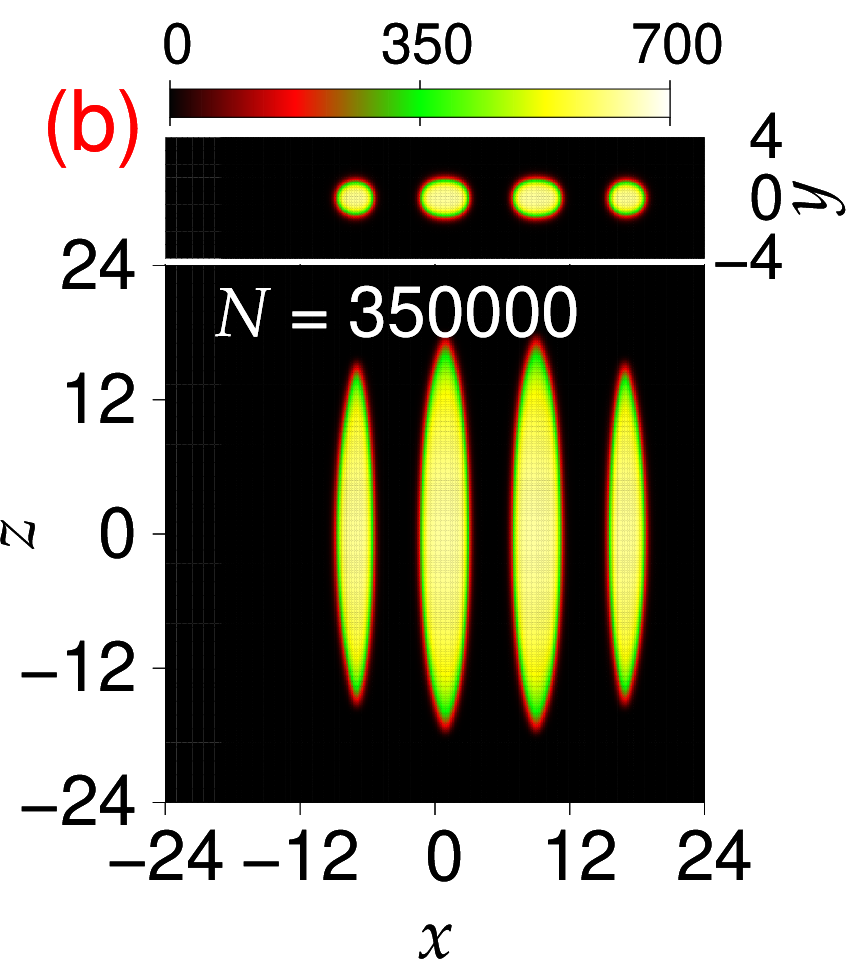}
\includegraphics[width=.32\linewidth]{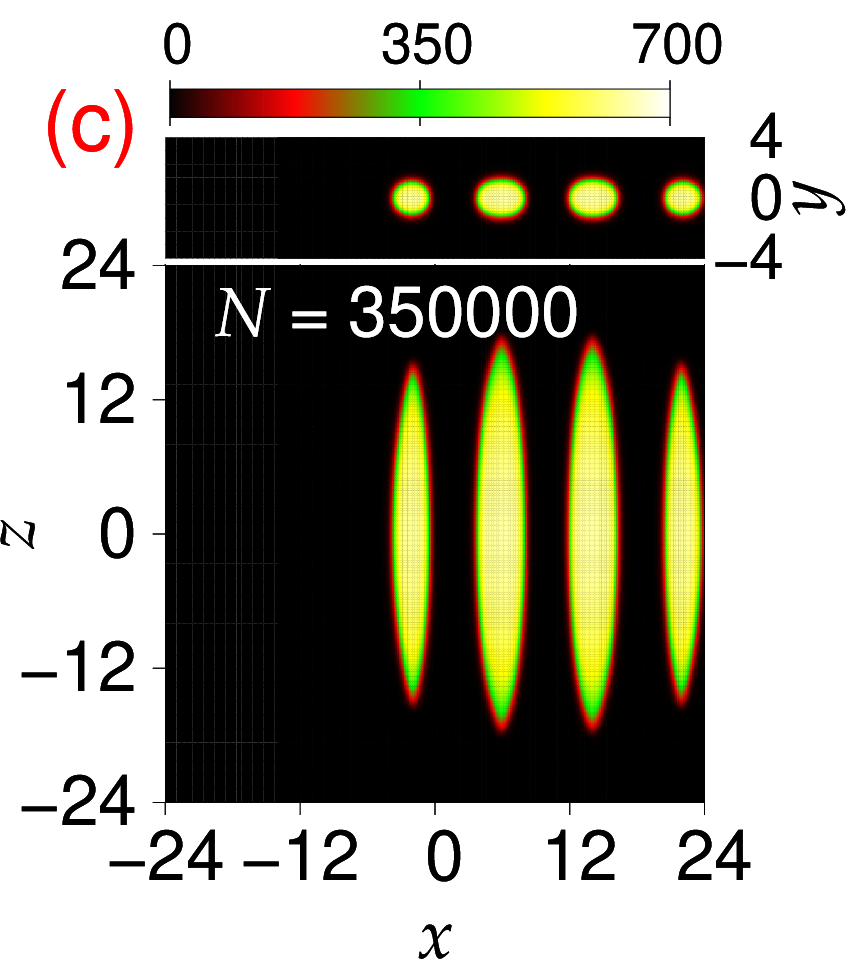}
\includegraphics[width=.32\linewidth]{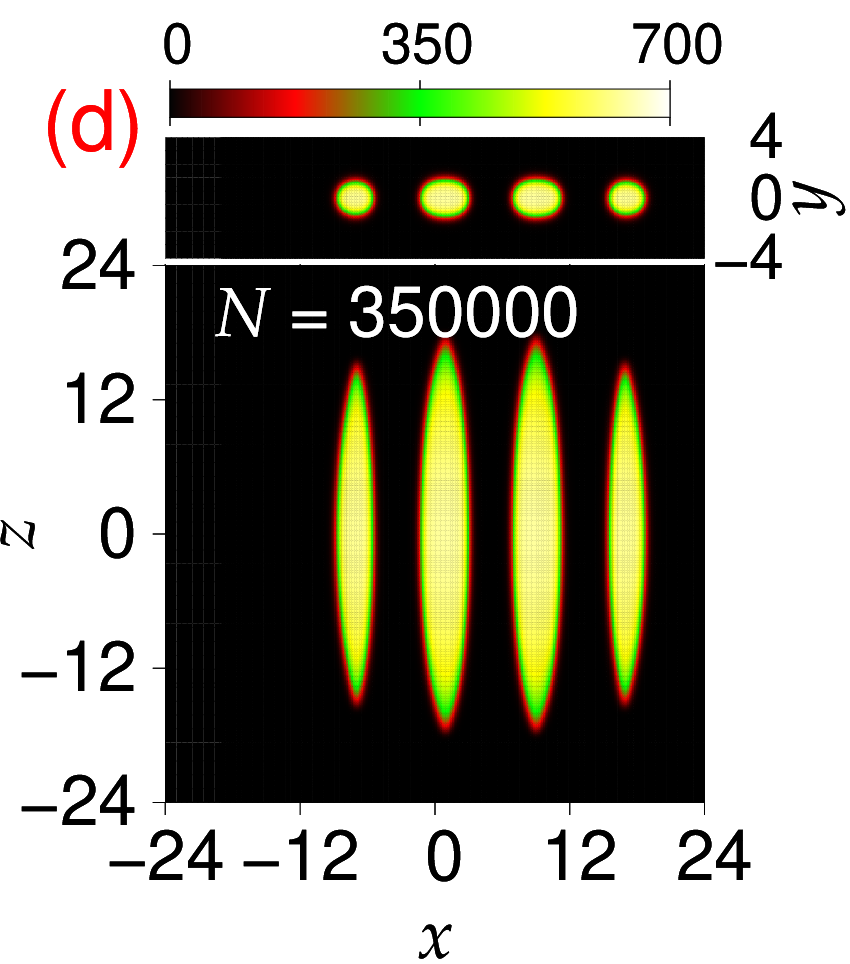}
\includegraphics[width=.32\linewidth]{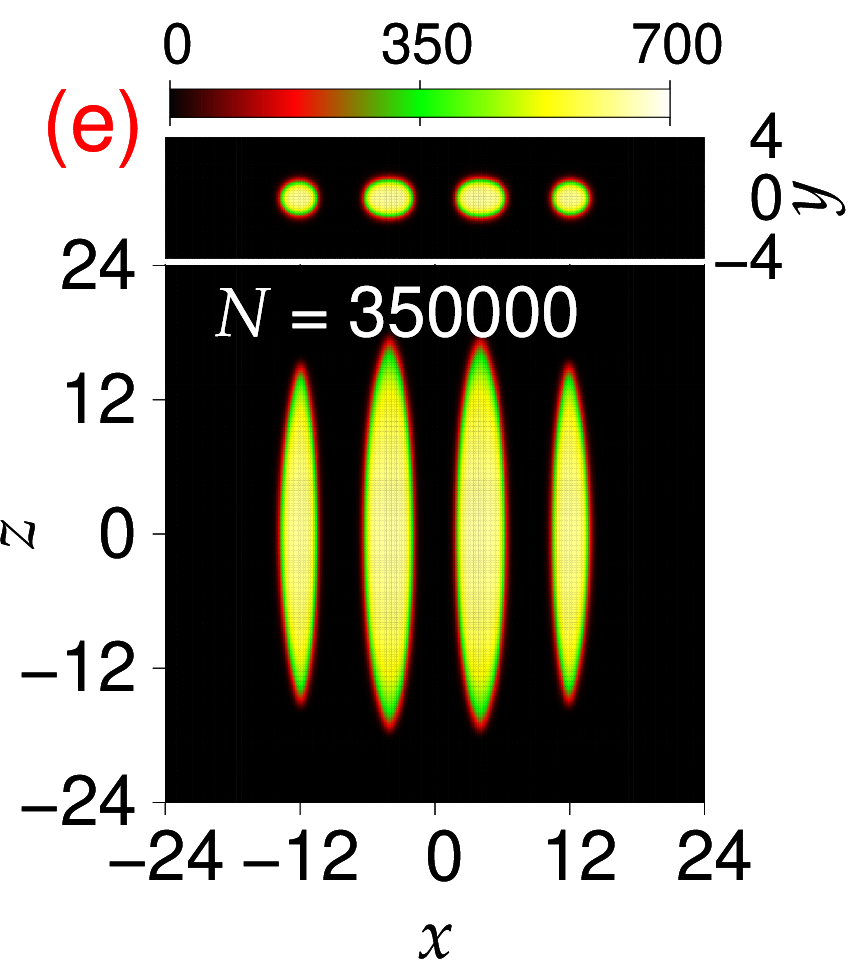}
\includegraphics[width=.32\linewidth]{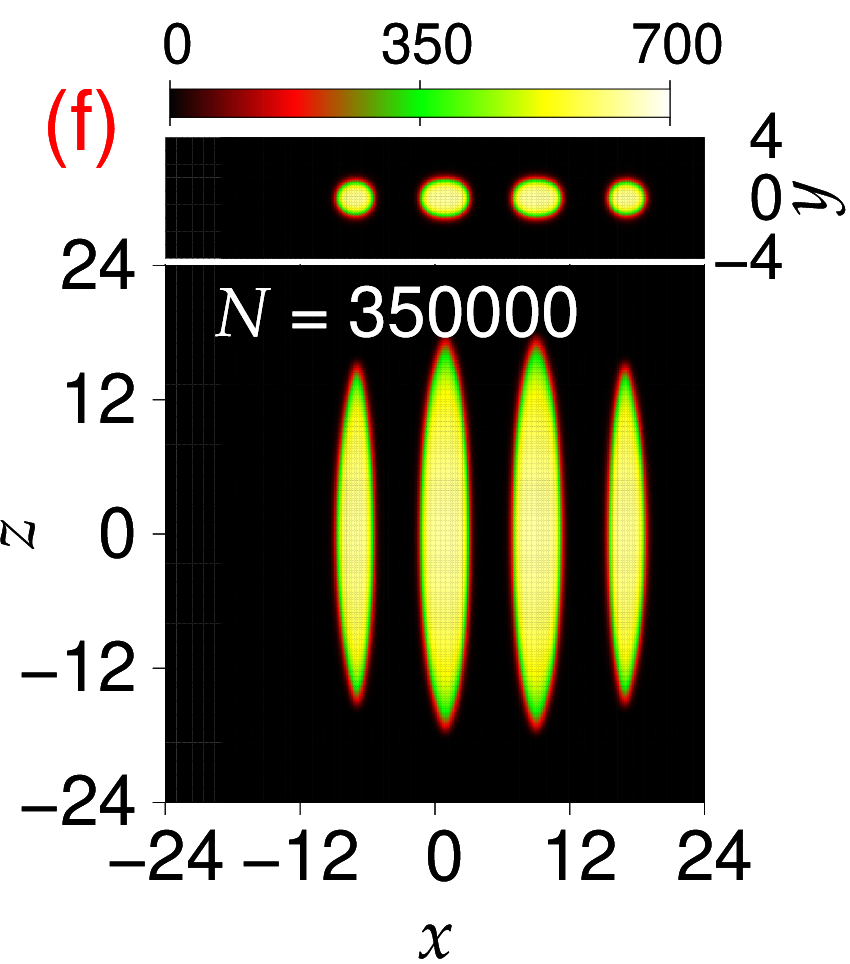}

\caption{ Contour plot of density $N|\psi(x,y,0)|^2$ (upper panel) and $N|\psi(x,0,z)|^2$ (lower panel)
of the quasi-1D four-droplet supersolid  of $N=350000$ $^{164}$Dy atoms of Fig. \ref{fig2}(h)
executing  {dipole-mode}  oscillation 
at times (a)
$t=0$, (b) $t=7.95$, (c) $t=15.9$, (d) $t=23.85$, (e) $t=31.8$, (f) $t=39.75$.  Other parameters are  $f_x=33/167$, $f_y=1, f_z = 60/167 $, $a=85a_0/l, a_{\mathrm{dd}}=130.8a_0/l$.
 }
\label{fig6} 
\end{center}
\end{figure}

 The dipole-mode oscillation in the $x$ direction is better illustrated
by snapshots of contour plot of density $N|\psi(x, y, 0)|^2$  and  $N|\psi(x, 0, z)|^2 $
in Fig. \ref{fig6}   at  times (a) $t = 0$, (b) $t = 7.95$,
(c) $t = 15.9$, (d) $t = 23.85$, (e) $t = 31.8$ and (f) $t = 39.75$. The supersolid starts
the oscillation in (a), passes through the position of the
minimum of trapping potential at $x = 5$ in (b) at $t = 7.95$
to the position of maximum displacement $x = 10$ in (c) at
$t = 15.9$. Then it turns around, passes again through the 
position $x = 5$  in (d) at $t = 23.85$ to the initial position $x = 0$
in (e) after a complete oscillation at $t = 31.8$, and repeats the same dynamics.  The theoretical period of oscillation $2\pi/\omega_x= 2\pi/(33/167)= 31.7967$ \cite{stg} compares well with the numerical value of 31.8.
There is no visible change 
of the crystalline structure during oscillation $-$ the four-droplet quasi-1D supersolid executes  linear oscillation along the $x$ direction like a rigid body.
 Sustained 
dipole-mode oscillation without distortion of 
the supersolid guarantee  both the 
superfluidity and the robustness of the crystalline structure.

\begin{figure}[t!]
\begin{center}
\includegraphics[width=.49\linewidth]{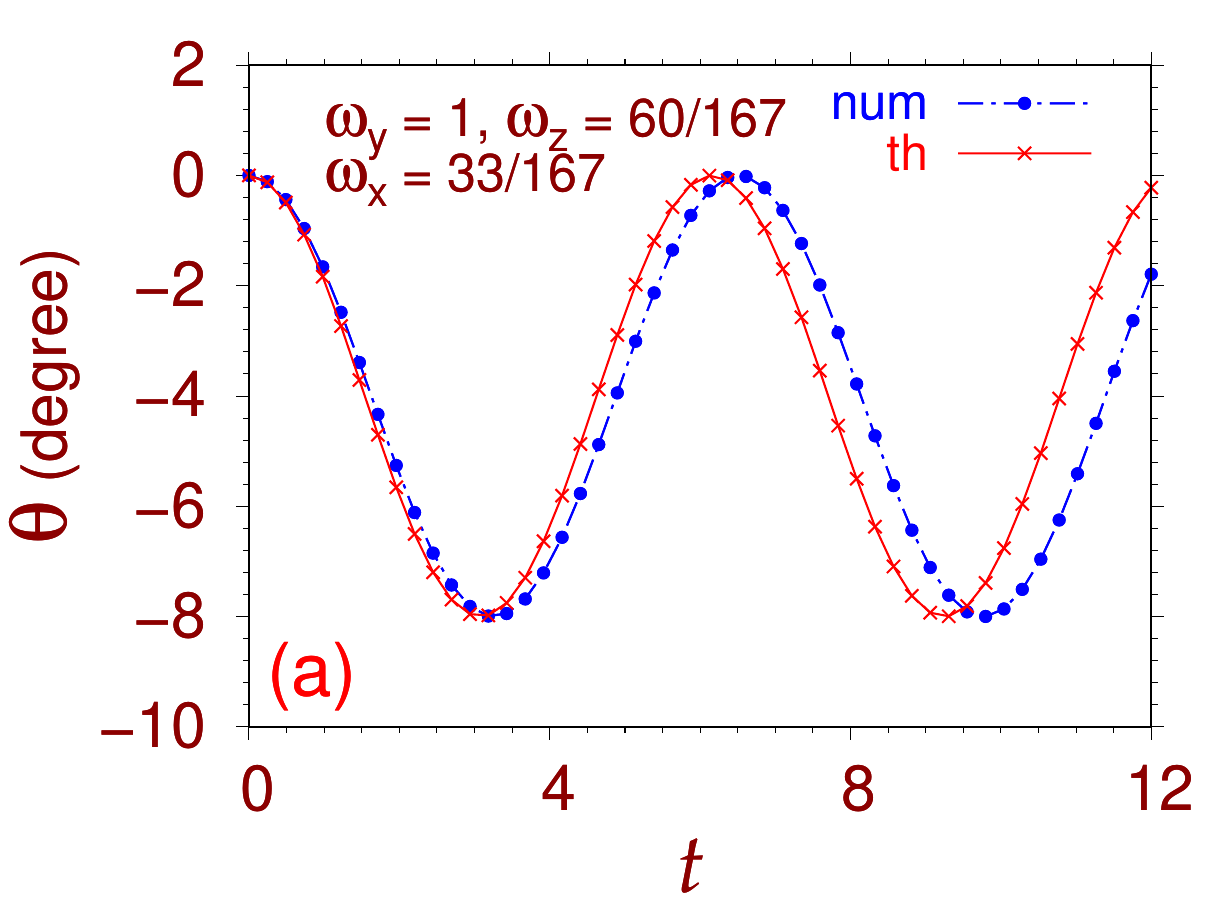}
\includegraphics[width=.49\linewidth]{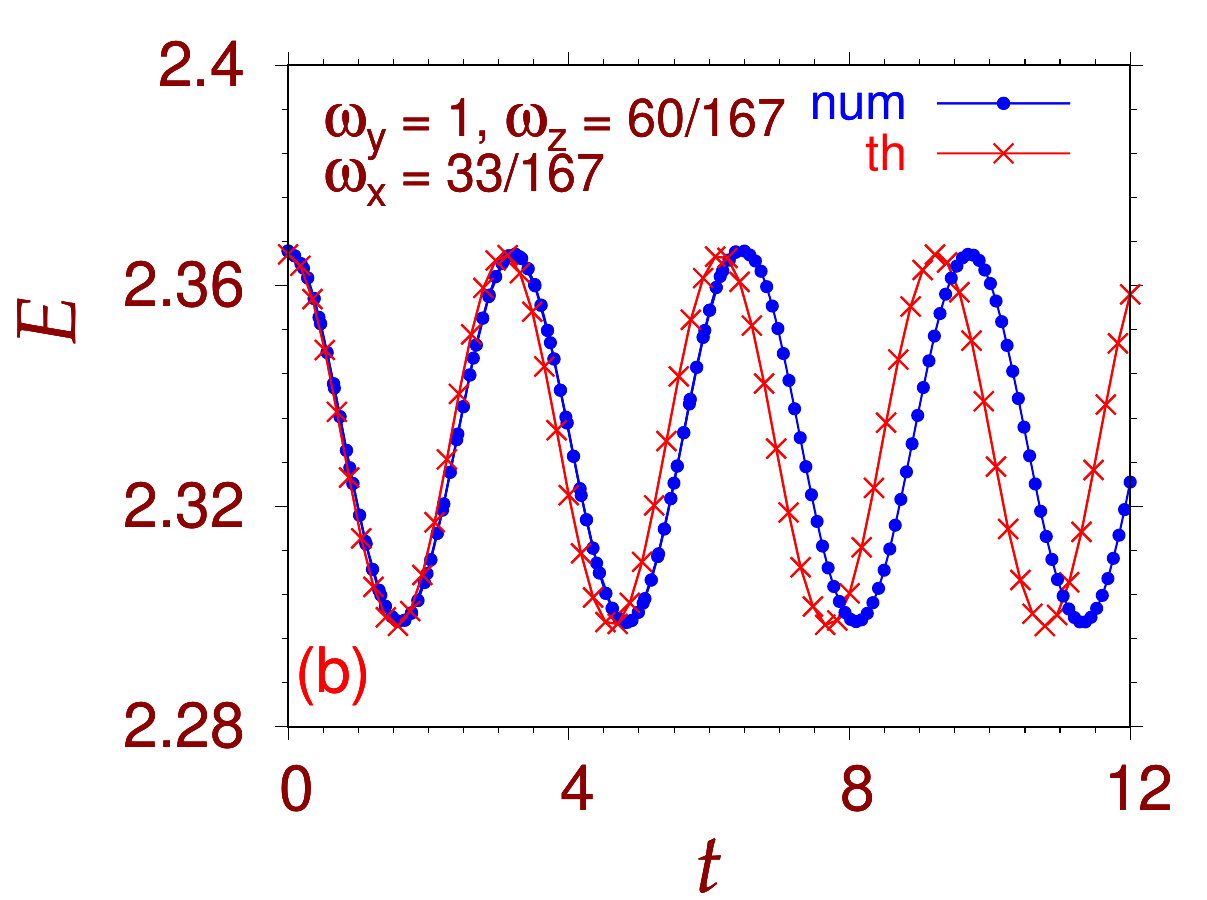}
\caption{(a) Numerical angular displacement $\theta$ (num) versus dimensionless time $t$ of a  quasi-1D three-droplet 
 supersolid of $N=150000$ $^{164}$Dy atoms  executing scissors-mode oscillation fitted to its theoretical estimate  $\theta \equiv -4+4\cos(\omega_{\mathrm{th}}t) =  -4+4\cos(1.0193369t)$ (th) .  
%
%$\cos(0.935\omega_{\mathrm{th}} t)$  $(\cos(0.971\omega_{\mathrm{th}} t))$  \cite{scith,sciex2}.  
% The parameters in trap (B) (A)  are $\omega_x = 23/90$ (33/167),  $\omega_y = \omega_z=1$, $ \omega_{\mathrm{th}}=\sqrt{\omega_x^2+\omega_y^2} =1.032138$ (), $a=
% 85a_0/l$, and $l=0.8275$. 
 The oscillation is started by giving a rotation of $\theta_0 =-4\degree$ of the trapping potential at $t=0$, viz. Eq. (\ref{scitrap}). (b) Numerical energy $E$ (num) versus time $t$  during the scissors-mode oscillation compared to its theoretical estimate (th). Other parameters are  $f_x=33/167$, $f_y=1, f_z = 60/167 $, $a=85a_0/l, a_{\mathrm{dd}}=130.8a_0/l$.
 }
\label{fig7}   
\end{center}
\end{figure}

\begin{figure}[t!]
\begin{center}
\includegraphics[width=.32\linewidth]{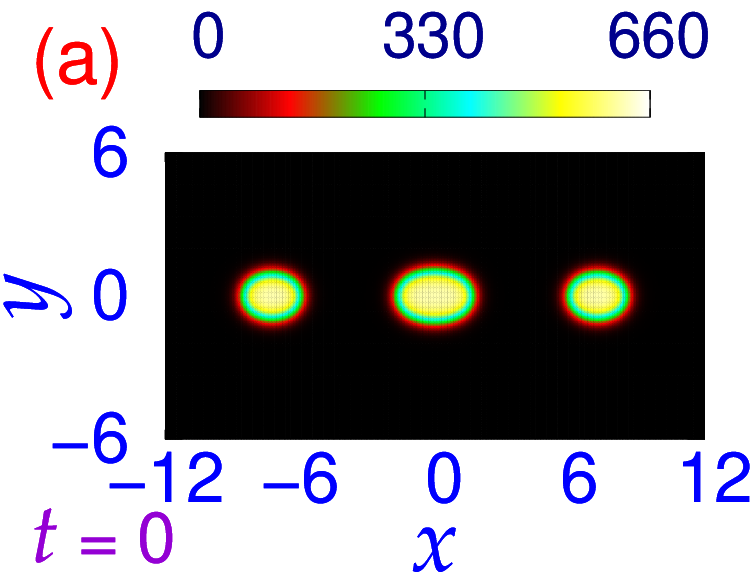}
\includegraphics[width=.32\linewidth]{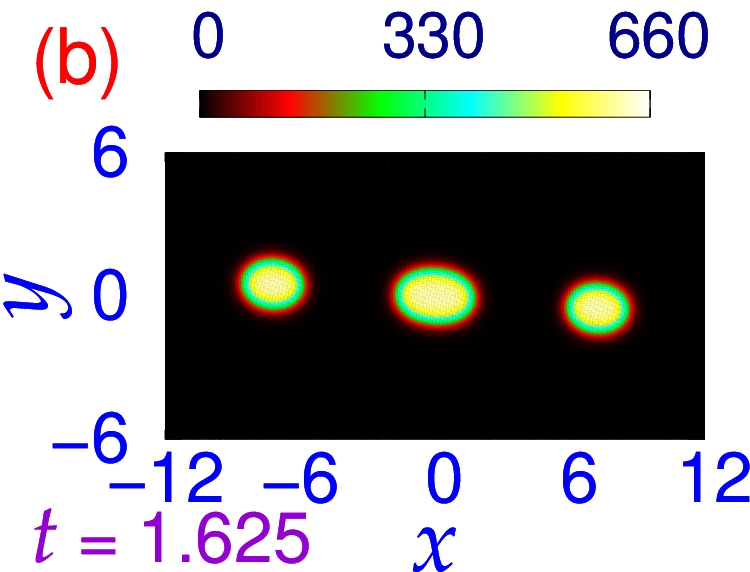}
\includegraphics[width=.32\linewidth]{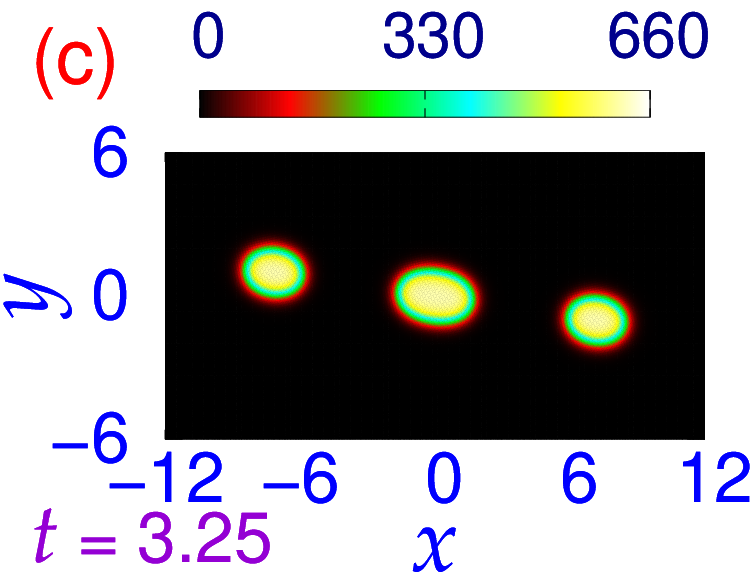}
\includegraphics[width=.32\linewidth]{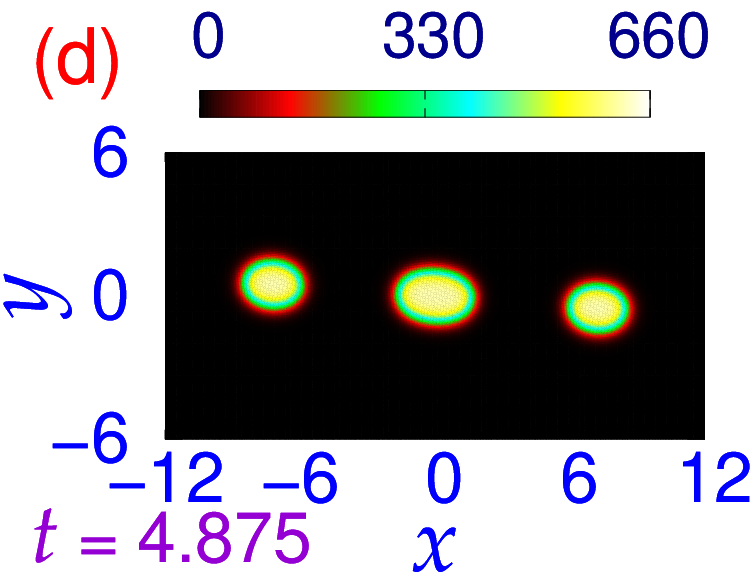}
\includegraphics[width=.32\linewidth]{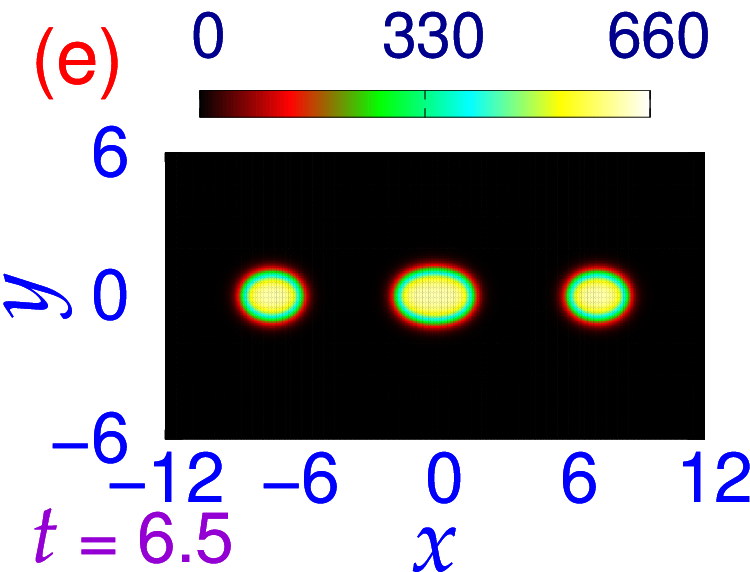} 
\includegraphics[width=.32\linewidth]{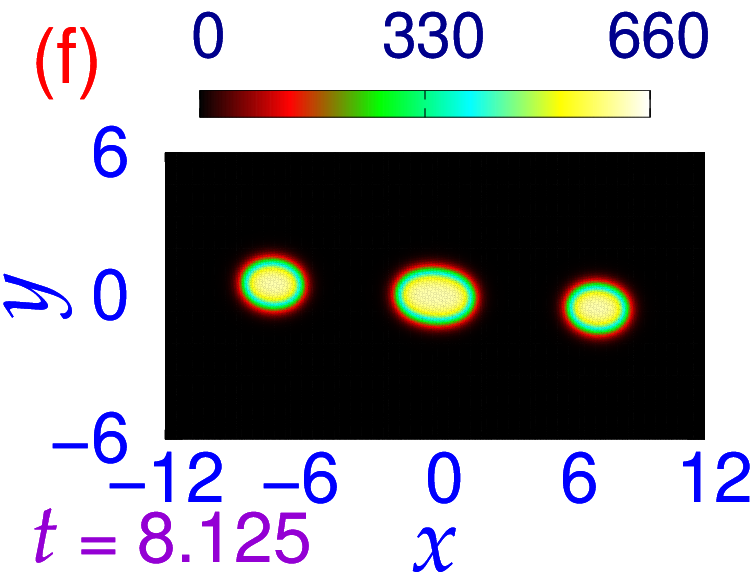} 
 
\caption{  Contour plot of density $N|\psi(x, y, 0)|^
2$  of the quasi-1D
three-droplet supersolid of $N =150000$ $^{164}$Dy atoms  of
Fig. \ref{fig2}(e) executing scissors-mode oscillation at times (a) $t = 0$ ($\theta =0$), (b) $t =
1.625$ ($\theta =-4\degree$), (c) $t = 3.25$ ($\theta =-8\degree$), (d) $t = 4.875$ ($\theta =-4\degree$), (e) $t = 6.5$ ($\theta =0$), and (f) $t = 8.125$ ($\theta =-4\degree$).   Other parameters are  $f_x=33/167$, $f_y=1, f_z = 60/167 $, $a=85a_0/l, a_{\mathrm{dd}}=130.8a_0/l$.
}
\label{fig8} 
\end{center} 

\end{figure}

 To further establish the supersolidity of the spatially-periodic states, we study  the scissors-mode oscillation of the quasi-1D three-droplet supersolid of Fig. \ref{fig2}(e) initiated by a rotation of the 
 spatially-asymmetric trap in the $x$-$y$ plane through an angle $\theta_0=-4\degree$   at $t=0$, viz. (\ref{scitrap}), by real-time propagation using the converged imaginary-time stationary state wave function of the ground state for $N=150000$.    Due to the strong spatial asymmetry ($\omega_x=33/167, \omega_y=1$) of the trap in the
$x$-$y$ plane the dipolar supersolid will execute sustained
scissors-mode oscillation \cite{scith,sci-y} around the $z$ direction with the angular frequency $\omega_{\mathrm{th}} \equiv \sqrt{(\omega_x^2+\omega_y^2)} =1.0193369$, which corresponds to the period 
$2 \pi/\omega = 6.16399.$ The numerical period of oscillation is approximately 6.5.
 Both the rotation angle $\theta$ and the 
 energy $E$ of the oscillating supersolid are found to execute a steady sinusoidal oscillation
as shown in Figs. \ref{fig7}(a)-(b), where we compare the numerical results of these oscillations with the respective theoretical estimates \cite{scith}. The frequency of the energy oscillation is double that of the angular oscillation. The time-evolution of the angular oscillation is more clearly demonstrated in Fig. \ref{fig8} through a snapshot of 
 density $N|\psi(x,y,0)|^2$ at different times (a) $t = 0$, (b) $t = 1.625$,
(c) $t = 3.25$, (d) $t = 4.875$, (e) $t = 6.5$, and (f) $t = 8.125$. At $t=0$ the supersolid with three droplets lies along the $x$ axis in its initial  position in (a). At $t=1.625$ it has rotated through an angle of $4\degree$ to the position of the rotated trap at $\theta =-4\degree$ in (b) corresponding to a minimum of energy.   At $t=3.25$ it has rotated through an angle of $8\degree$ to the position of maximum displacement  $\theta =-8\degree$ in (c) corresponding to a maximum of energy. After that at   $t=4.875$ the supersolid turns around and comes to the minimum-energy position in (d). Finally, at $t=6.5$  the supersolid comes to its initial position $\theta=0$ at the end of a complete cycle of oscillation in (e). After that the same periodic oscillation is repeated. A quasi-2D supersolid formed in a quasi-2D trap usually does not execute a prolonged scissors-mode oscillation \cite{dyna,related}. But the present quasi-1D supersolid  formed in a quasi-2D trap is
 demonstrated to execute sustained scissors-mode oscillation.  However, the frequency of this scissors-mode oscillation is smaller than its theoretical estimate \cite{scith,related}.

\section{Summary}

\label{IV}
  
We demonstrate the formation of a new type of droplet and a new type of crystallization of these droplets in a quasi-1D supersolid in a strongly  dipolar BEC confined by a quasi-2D trap in $x$-$z$ plane, where $z$ is the polarization direction of dipolar atoms.  This confinement is caused by a strong trap in the $y$ direction, while in other studies of dipolar supersolids a strong trap is applied in the polarization $z$ direction.  The new scenario corresponds to a swapping of the trap frequencies  art.zip  $\omega_y \leftrightarrow \omega_z$ maintaining 
  the same overall trapping 
$\bar \omega = \sqrt[3]{\omega_x \omega_y\omega_z}.$
In the new scenario a single droplet can be formed for a relatively small number $-$ about 1000 $-$   of $^{164}$Dy atoms, whereas in previous studies \cite{2d2exFer,2d4,luis1} a single droplet can be formed with a large number $- (\gtrapprox 20000) -$ of atoms.   The present single droplet is called a mini droplet. 
As the number of atoms is increased, in the new scenario a quasi-1D supersolid is formed  in the quasi-2D trap   with a small number of droplets arranged in a spatially-periodic lattice along the $x$ axis, viz. Fig. \ref{fig2},  each of these droplets containing a large number $-$ between 50000 and 200000 $-$ of atoms.
  Such droplets are termed mega droplets. The quasi-1D chain of droplets along the $x$ axis appears as a stripe pattern in the $x$-$z$ plane.
  In the previous studies \cite{2d2exFer,2d4,blakieprl,luis1}, viz. Fig. \ref{fig3}, in the quasi-2D trap a quasi-2D supersolid  with a large number of droplets was formed, each of these droplets containing a small number of atoms.

In addition to the study of density of the stationary ground and a few metastable states by imaginary-time propagation we also studied the dipole-mode and scissors-mode oscillations of the quasi-1D supersolid 
in order to establish the rigidity of the crystalline structure and the superfluidity of the supersolid by  real-time propagation using the converged imaginary-time wave function as the initial state. The dipole-mode oscillation was initiated by displacing the trap along the $x$ axis through a small distance at $t=0$. The scissors-mode oscillation was initiated by giving a rotation of the trap through a small angle around the $z$ axis  at $t=0$.  In both cases a continued steady linear and angular harmonic oscillation  was obtained and this demonstrates the  supersolidity of the system.  The findings of this study can be verified with the present experimental set up of Ref.  \cite{2d2exFer}, as the present set up corresponds to only a swapping of the trap frequencies $\omega_y \leftrightarrow  \omega_z$ used in that experiment.

\section*{CRediT authorship contribution statement}
L. E. Young-S. and S. K. Adhikari: Conceptualization, Methodology, Validation, Investigation, Writing – original draft, Writing – review and  editing,
Visualization.

\section*{Declaration of competing interest}
The authors declare that they have no known competing financial interests or personal relationships that could have appeared
to influence the work reported in this paper.
\section*{Data availability}
No data was used for the research described in the article.

\section*{Acknowledgments}
SKA
 acknowledges support by the CNPq (Brazil) grant 301324/2019-0. The use of the supercomputing
cluster of the Universidad de Cartagena, Cartagena, Colombia  is gratefully acknowledged.

%\end{acknowledgments}

\end{document}